\documentclass[11pt]{article}
\usepackage{amsmath}
\usepackage{amssymb}
\usepackage{amsthm}
\usepackage{amsfonts}
\usepackage{comment}
\usepackage{color}
\usepackage{mathrsfs}
\usepackage{braket}
\usepackage{graphicx}
\usepackage[subrefformat=parens]{subcaption}
\usepackage{cite}
\usepackage{here}
\usepackage[stable]{footmisc}

\makeatletter

\@addtoreset{equation}{section}
\makeatother

\begin{document}
\pagestyle{empty}

\title{\bf{Bound on Lyapunov exponent\\in Kerr-Newman-de Sitter black holes\\by a charged particle}}

\date{}
\maketitle

\begin{center}

{\large 
Junsu Park$^\star$\footnote{wnstn@dgu.ac.kr}
and Bogeun Gwak$^\star$\footnote{rasenis@dgu.ac.kr}
} \\
\vspace*{0.5cm}

{\it 
$^\star$Division of Physics and Semiconductor Science, Dongguk University, Seoul 04620, Republic of Korea
}

\end{center}

\vspace*{1.0cm}

\begin{abstract}
{\noindent
We investigate the bound on the Lyapunov exponents by a charged particle in Kerr-Newman-de Sitter black holes using analytic and numerical methods. We determine whether the Lyapunov exponent can exceed the bound by an electrically charged particle with an angular momentum. Our tests are applied to the de Sitter spacetime by the positive cosmological constant such as Reissner-Nordstr\"{o}m-de Sitter, Kerr-de Sitter, and Kerr-Newman-de Sitter black holes. In particular, we consider Nariai and ultracold limits on these black holes for our tests. From our analysis results, there remain violations on the bound under the positive cosmological constant, and electric charge and angular momentum of the particle significantly impact the Lyapunov exponent.

}
\end{abstract}

\newpage
\baselineskip=18pt
\setcounter{page}{2}
\pagestyle{plain}
\baselineskip=18pt
\pagestyle{plain}

\setcounter{footnote}{0}

    \section{Introduction}

    Black holes can be considered as solutions of the Einstein field equation and have the singularity and event horizon. The event horizon has a spherical shape. Within the event horizon, the spacetime is curved owing to gravity, so there is no outgoing geodesic to go outside the black hole crossing the event horizon. An ingoing geodesic goes to the endpoint in the black hole, which is the singularity. Thus, we can find properties of the geometry of black holes by probing the geodesic. Therefore, an observer outside the black hole cannot observe energy and matter that radiates from inside the black hole.

    When considering quantum effects, we find that black holes radiate energy, which is Hawking radiation \cite{Hawking:1975vcx, Hawking:1976de}. This radiation implies that the black hole can be described as a thermal system having a temperature. This is the Hawking temperature, and is proportional to the surface gravity of a black hole. Further, black holes have an entropy that is proportional to its surface area, and which is called the Bekenstein-Hawking entropy \cite{Bekenstein:1973ur, Bekenstein:1974ax}. In addition, the cosmological constant affects the metric of the black hole, and is related to the asymptotic boundary of the spacetime. When a black hole has a negative cosmological constant, we call it an anti-de Sitter (AdS) black hole.

    Recently, a relationship has been discovered between the AdS spacetimes and the quantum field theory, and is called the anti-de Sitter/conformal field theory (AdS/CFT) correspondence \cite{Maldacena:1997re, Gubser_1998, witten1998anti}. The AdS/CFT correspondence explains the duality between gravity and quantum field theory. The CFT is quantum field theory, which is invariant under conformal transformations. According to the AdS/CFT correspondence, the (d+1)-dimensional gravity theory with a negative cosmological constant is associated with a d-dimensional CFT on the boundary of AdS spacetimes. For CFT at a zero temperature, it is dual to an AdS spacetime with no black hole, and CFT with a finite temperature is dual to an AdS black hole \cite{Witten:1998zw}.

   When the system is sensitive to initial conditions, it is chaotic. We can measure the magnitude of chaos using the Lyapunov exponent, which measures how the difference in initial conditions affects the subsequent motion. If the Lyapunov exponent of the motion is positive, the motion is chaotic. In a thermal quantum system, there is another indicator that measures the magnitude of chaos, which are out-of-time order correlators (OTOCs) \cite{Maldacena:2015waa}. Its definition is the thermal expectation value of the commutator of two Hermitian operators at a finite temperature. Using OTOCs, we can also obtain the Lyapunov exponent in a thermal quantum system. Recently, by using the AdS/CFT correspondence, it was assumed by Maldacena, Shenker, and Stanford\cite{Maldacena:2015waa} that the Lyapunov exponent has an upper bound in a thermal quantum system. The upper bound is associated with the temperature of the thermal system.

     It is known that the geodesic in a black hole can be chaotic, and this black hole chaos has been researched in \cite{Dettmann:1994dj, Suzuki:1996gm, Suzuki:1999si, Han:2008zzf, Verhaaren:2009md, Pradhan:2012rkk, Bombelli:1991eg, dettmann1995chaos, Barrow:1981sx, Pradhan:2012qf, Pradhan:2013bli, Chen:2016tmr, Dalui:2018qqv, Li:2018wtz, Cubrovic:2019qee, Giataganas:2021ghs}. According to Hashimoto and Tanahashi \cite{Hashimoto:2016dfz}, it is assumed that the Lyapunov exponent of a particle in a black hole has an upper bound near the horizon. When there is an extremal force on a particle, the motion of the particle can be described in the inverse harmonic oscillator, and we can obtain the Lyapunov exponent of the particle. The maximum value of the Lyapunov exponent of the particle in a black hole satisfies the conjecture. The bound is associated with the Hawking temperature of the black hole instead of the temperature of the thermal quantum system \cite{Hashimoto:2016dfz}. The bound has been tested many times by varying the geometry of the spacetime in \cite{Dalui_2019, Colangelo_2022, _ubrovi__2019, Zhao:2018wkl, Lei:2020clg, Kan:2021blg, Yu:2022ysm, Gwak:2022cha, Jeong:2023hom}.
     
     In this work, we investigate the bound on the Lyapunov exponent in Kerr-Newman-dS (KNdS) black holes by a charged particle with angular momentum. By testing the bound by a particle with angular momentum, we find differences in the values of the Lyapunov exponents and the validity on the bound, which was originally proposed on a particle with the radial momentum only\cite{Hashimoto:2016dfz}. In particular, compared with cases of zero and negative cosmological constants\cite{Kan:2021blg,Gwak:2022cha}, KNdS black holes have two more extremal cases: Nariai and ultracold black holes. Hence, the Lyapunov exponents on these black holes were also tested and their relationship to the bound clarified. Here, we study the Lyapunov exponents by performing analytical and numerical approaches on Reissner–Nordstr\"{o}m-dS (RNdS), Kerr-dS (KdS), and KNdS black holes. Owing to the complicated calculations required, most of the Lyapunov exponents are obtained by performing extensive numerical computations, but our analysis successfully provides a general perspective of Lyapunov exponents in dS spacetimes. Furthermore, we finally complete testing the bound on the Lyapunov exponent using an arbitrary cosmological constant\cite{Kan:2021blg,Gwak:2022cha}.  
     
     Before we calculate the Lyapunov exponent, we briefly review the KNdS black hole in Section 2. Then, we calculate the Lyapunov exponent of a test particle in the KNdS black hole and check the bound analytically with a low spin-charge limit, near-horizon limit, and near cosmological horizon limit in Section 3. In Sections 4 and 5, we numerically calculate the Lyapunov exponent of particles in RNdS black holes and KdS black holes, respectively. In Section 6, we calculate the Lyapunov exponent in KNdS black holes. Finally, we summarize our results in Section 7.

    \section{Review on KNdS black holes}
    
The KNdS metric is a stationary solution of the Einstein-Maxwell equation.
The KNdS black hole has a rotating charged mass. In addition, the positive cosmological constant affects the geometry of the spacetime. The KNdS metric is described in the Boyer-Lindquist coordinates as \cite{Carter:1968ks}:
    \begin{align}
    \label{eq:Kerr-Newman dS black hole}
		ds^2&=-\frac{\Delta_r}{\rho^2}\left(dt-\frac{a\sin^2\theta}{\Xi}d\phi\right)^2+\frac{\rho^2}{\Delta_r}dr^2+\frac{\rho^2}{\Delta_\theta}d\theta^2+\frac{\Delta_\theta\sin^2\theta}{\rho^2}\left(adt-\frac{r^2+a^2}{\Xi}d\phi\right)^2 \\
		&=-\frac{1}{\rho^2}\left(\Delta_r-a^2\Delta_\theta\sin^2\theta\right)dt^2+\frac{2a\sin^2\theta}{\Xi \rho^2}\left(\Delta_r-(r^2+a^2)\Delta_\theta\right)dtd\phi \nonumber \\
		&\qquad +\frac{\sin^2\theta}{\Xi^2\rho^2}\left((r^2+a^2)^2\Delta_\theta-a^2\Delta_r\sin^2\theta\right)d\phi^2+\frac{\rho^2}{\Delta_r	}dr^2+\frac{\rho^2}{\Delta_\theta}d\theta^2,
    \end{align}
where 
    \begin{gather}
		\rho^2=r^2+a^2\cos^2\theta, \\
		\Xi=1+\frac{a^2}{3}\Lambda \label{eq:Xi} \\
		\Delta_r=(r^2+a^2)\left(1-\frac{r^2}{3}\Lambda\right)-2Mr+Q^2, \\
		\Delta_\theta=1+\frac{a^2}{3}\Lambda\cos^2\theta.
    \end{gather}
  $M$, $Q$, and $a$ are the mass, electric charge, and spin parameters, respectively. $\Lambda$ is the positive cosmological constant. We only consider the electric charge in the metric (\ref{eq:Kerr-Newman dS black hole}).
 The electromagnetic potential is 
    \begin{align}
        A=-\frac{Qr}{\rho^2}\left(dt-\frac{a\sin^2\theta}{\Xi}d\phi\right).
        \label{eq:EM_potential}
    \end{align}
  The electric charge, angular momentum, and mass of the KNdS black hole are respectively given as \cite{Caldarelli_1999}
    \begin{align}
      Q_B = \frac{Q}{\Xi^2} ,\quad  J_B = \frac{Ma}{\Xi^2},  \quad M_B = \frac{M}{\Xi^2}.
    \end{align}
The radius of horizons can be obtained by solving $\Delta_r = 0$. We can obtain four real roots when the below condition is satisfied.

\begin{figure}[ht]
    \centering
        \begin{subfigure}{0.4\textwidth}
            \centering 
            \includegraphics[width=1\linewidth]{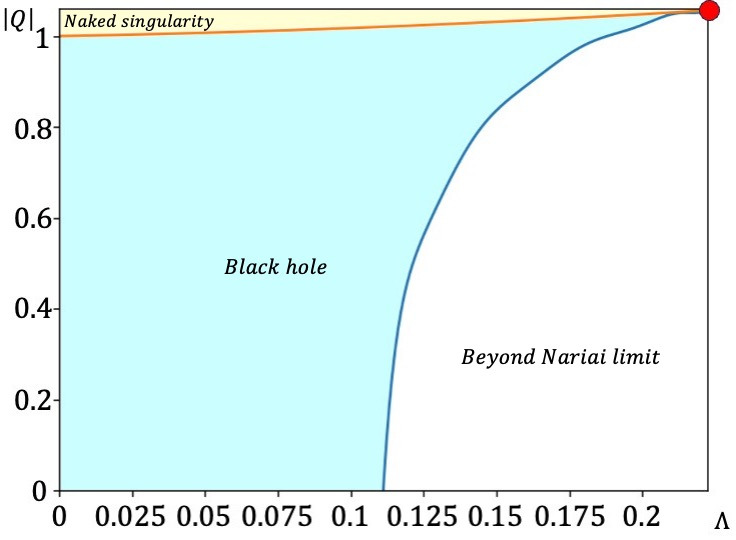}
            \caption{RNdS black hole}
        \end{subfigure}
        \begin{subfigure}{0.4\textwidth}
            \centering
            \includegraphics[width=1\linewidth]{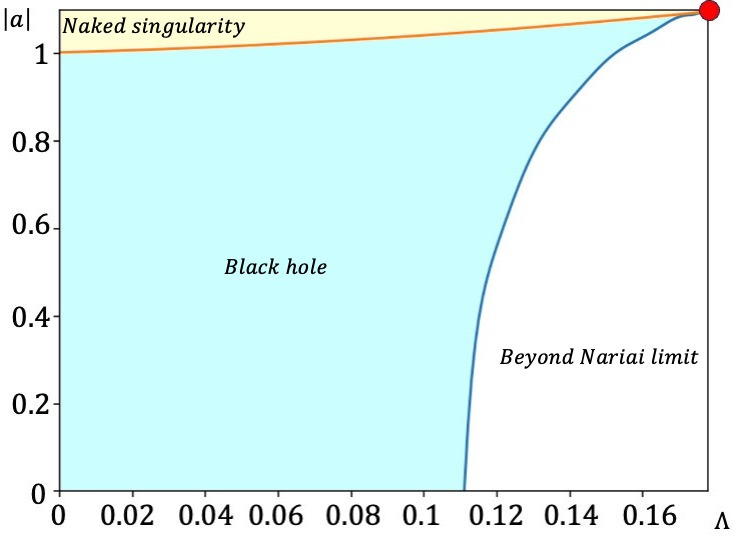}
            \caption{Kerr-dS black hole}
        \end{subfigure}
        \begin{subfigure}{0.4\textwidth}
            \centering
            \includegraphics[width=1\linewidth]{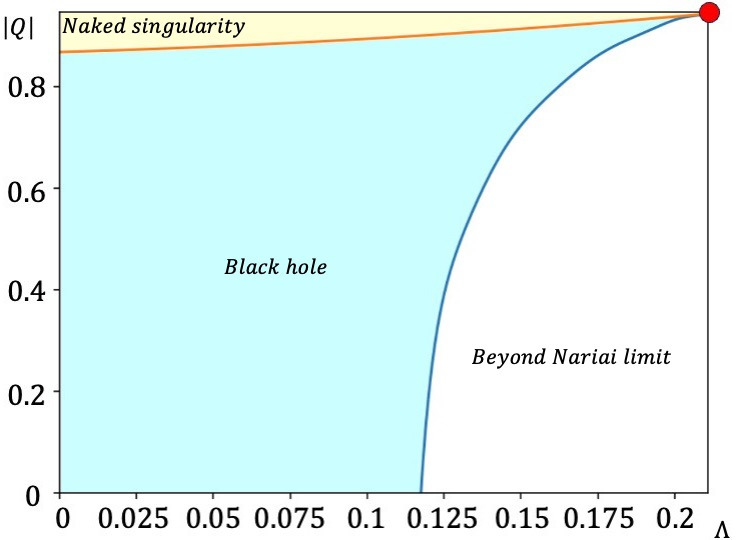}
            \caption{KNdS black hole ($a = 0.5$)}
        \end{subfigure}
        \begin{subfigure}{0.4\textwidth}
            \centering
            \includegraphics[width=1\linewidth]{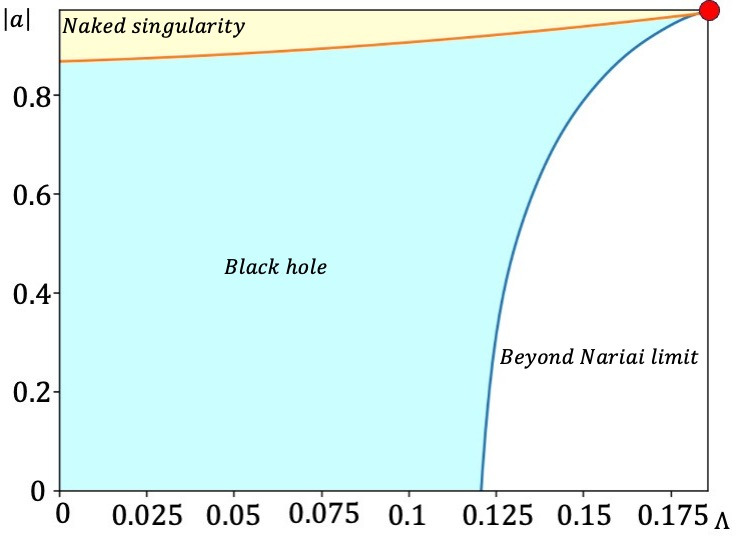}
            \caption{KNdS black hole ($Q = 0.5$)}
        \end{subfigure}
    \caption{Plot of graphs describing the region where the black hole exists. (a), (b), (c), and (d) are graphs for the RNdS black hole, KdS black hole, KNdS black hole with $a = 0.5$, and KNdS black hole with $Q = 0.5$, respectively. The vertical axis is the absolute value of the electric charge or spin parameter of the metric. The horizontal axis is the cosmological constant. On the orange line, the black hole exists as the extremal black hole. On the blue line, the black hole exists as the Nariai black hole. These two lines meet at a red point to the upper right. At the red point, the black hole exists as the ultracold black hole. In the sky-blue region, a black hole can physically exist, while in the yellow region, the black hole becomes the naked singularity. The white region is beyond the Nariai limit.}
    \label{fig:blackhole_region}
    \end{figure}

 We can obtain four real roots when the below condition is satisfied as
     \begin{align}
        \frac{1}{9}\Lambda^2 + a^4 - \frac{14}{3}a^2 \Lambda - 4Q^2\Lambda \ge 0
        \label{eq:condition1}
    \end{align}
When the condition (\ref{eq:condition1}) is satisfied, we obtain a negative root $r_{--}$ and three positive roots, $r_- \le r_+ \le r_{c}$ \cite{Dehghani:2003uu}. $r_-$ is an inner horizon, $r_+$ is an outer horizon, and $r_{c}$ is a cosmological horizon. For special cases, two or three horizons can be located at the same radius. For these cases, the black hole has been named the extremal, Nariai, or ultracold black hole. They are often mentioned as the extremal black hole or extremal case at once. The extremal black hole has the same inner and outer horizon, $r_- = r_+$, while the Nariai black hole has the same outer horizon $r_+$ and cosmological horizon $r_{c}$ \cite{Narai:1999.11 , Narai:1999.22}. Finally, the ultracold black hole has all the same horizons, $r_- = r_+ = r_{c}$. These extremal cases have zero surface gravity at the event horizon $r = r_+$. Because the cosmological horizon $r_{c}$ only exists in the dS black hole, the Nariai and ultracold black holes are the extremal cases only for the dS black hole. The parameters of these three extremal cases satisfy the equation below \cite{Dehghani:2003uu}.
    \begin{align}
        &\Lambda^3M^4+(\frac{11}{3}\Lambda^2a^4 - \frac{4}{3}\Lambda^3Q^2-\frac{11}{9}\Lambda^3a^2+4\Lambda^2Q^2a^2+\frac{1}{3}\Lambda a^6 - \frac{1}{81}\Lambda^4)M^2 \nonumber \\
        &\quad +a^8Q^2 + \frac{22}{9}\Lambda^2a^4Q^2+ \frac{4}{9}\Lambda^3a^2Q^2+\frac{2}{3}a^6\Lambda^2 +\frac{4}{27}\Lambda^3a^4 + \frac{32}{9}\Lambda^2a^2Q^4 + \frac{1}{81}\Lambda^4Q^2 \nonumber \\
        &\quad + 4a^6\Lambda Q^2 + \frac{4}{3}\Lambda a^8 + \frac{1}{81}\Lambda^4a^2 + a^{10} + \frac{16}{9}\Lambda^2 Q^6 + \frac{8}{27}\Lambda^3 Q^4 + \frac{8}{3}\Lambda Q^4a^4 = 0
    \label{eq:criticalmass}
    \end{align}
 By solving the equation (\ref{eq:criticalmass}) for specific parameters, if we obtain two real roots, they are physical quantities for the Nariai and extremal black holes. The black hole whose parameter lies between these two roots can physically exist. If not, the black hole cannot physically exist. If we obtain a multiple root by solving the equation (\ref{eq:criticalmass}), it is a quantity for the ultracold black hole. Therefore, we can visualize the region in which the black hole physically exists. In Fig. \ref{fig:blackhole_region}, the graph describes these regions for the RNdS, KdS, and KNdS black holes.

It is necessary to check that the metric (\ref{eq:Kerr-Newman dS black hole}) is well-defined asymptotically, and this is done by introducing zero angular momentum observers (ZAMOs). The ZAMOs can be defined by their four velocity $u^\alpha$: $u_\alpha \phi^\alpha = 0$. The angular velocity of ZAMOs is given by
    \begin{align}
        \Omega = \frac{d\phi}{dt} &=-\frac{g_{t\phi}}{g_{\phi\phi}}.
    \end{align}
The angular velocity of ZAMOs is non-zero because the rotation of the black hole results in the dragging of inertial frames, which is called the frame-dragging effect. However, the frame-dragging effect will disappear when the observers are located at infinity.
The angular velocity of ZAMOs at infinity $\Omega_{\infty}$ is
    \begin{align}
        \Omega_{\infty} = \lim_{r \to \infty}\Omega  =\frac{a}{3}\Lambda.
    \end{align}
 The observer at infinity is not static owing to the cosmological constant $\Lambda$. We should use coordinate transformations to recover an asymptotically static observer\cite{Hawking:1998kw}.
        \begin{align}
		t\to T,\qquad \phi\to \Phi+\frac{1}{3}a\Lambda T.
        \label{eq:transformation}
	\end{align}
    Then, the observer has zero angular momentum at infinity, and the metric is well defined as asymptotically dS spacetime. Under the transformation (\ref{eq:transformation}), we can rewrite the metric as\cite{Gwak:2018akg,Gwak:2021tcl}
    \begin{align}
		ds^2&=-\frac{\Delta_r}{\rho^2\Xi^2}\left(\Delta_\theta dT-a\sin^2\theta d\Phi\right)^2+\frac{\rho^2}{\Delta_r}dr^2+\frac{\rho^2}{\Delta_\theta}d\theta^2+\frac{\Delta_\theta \sin^2\theta}{\rho^2\Xi^2}\left\{a\left(1-\frac{r^2}{3}\Lambda\right)dT-(r^2+a^2)d\Phi\right\}^2.
	\label{eq:KN_metric_static}
    \end{align}
 The vector potential (\ref{eq:EM_potential}) also becomes
    \begin{align}
		A=-\frac{Qr}{\rho^2\Xi}\left(\Delta_\theta dT-a\sin^2\theta d\Phi\right).
    \end{align}

Here, we briefly examine a conjecture about the chaos in a thermal quantum system and a bound in a black hole.
 In studying the chaos, the Lyapunov exponent $\lambda$ measures the sensitivity of the initial system, and is classically defined as
  \begin{align}
      \lambda = \lim_{t \to \infty}\lim_{\Delta x_0 \to 0}\frac{1}{t} \ln \left| \frac{\Delta x(t)}{\Delta x_0} \right|,
      \label{eq:Lyapunov}
  \end{align}
 where $\Delta x(t)$ and $\Delta x_0$ are the differences between two trajectories at arbitrary time $t$ and the initial time, respectively. When the Lyapunov exponent $\lambda$ is positive, we say the system is chaotic.

 In a thermal quantum system with a large degree of freedom, the OTOCs, $-\braket{[V(t), W(0)]}_\beta$, measure the magnitude of chaos, where $V(t)$ and $W(t)$ are any Hermitian operators and $\braket{\cdots}_\beta$ = $Z^{-1}tr[e^{-\beta H}]$ refers to the thermal expectation value at temperature $T = \beta^{-1}$. $Z$ is a partition function. The OTOCs can be written as \cite{Shenker:2013pqa, Shenker:2013yza,Roberts_2015,shenker2015stringy}
     \begin{align}
         F(t) = f_0 - \frac{f_1}{N^2} e^{2\pi t/\beta} + {\cal O}(N^{_4}),
         \label{eq:OTOC}
     \end{align}
where $f_0, f_1$ are positive constants depending on the operators V, W. There is an approximately constant factorized value $F_d = tr[y^{2}Vy^{2}V]tr[y^{2}W(t)y^{2}W(t)]$. If the system is chaotic, $F_{d} - F(t)$ increases exponentially, $F_{d} - F(t) \sim \exp \lambda_L t$ \cite{Larkin:1969qu, Arnoud:78.062329, Sekino:2008he, Kitaev:2014hi}. $\lambda_L$ is the Lyapunov exponent of the thermal quantum system. Then, the bound on chaos is estimated by using the Lyapunov exponent $\lambda_L$, which is given as \cite{Maldacena:2015waa}
    \begin{align}
          \lambda_L \le \frac{2 \pi T}{\hbar},
          \label{eq:bound on chaos}
    \end{align}
  where $T$ is the temperature of the thermal quantum system.
  The estimate (\ref{eq:bound on chaos}) assumes that the Lyapunov exponent cannot be larger than the product of the temperature of the system and the constants.
  
  It is considered that the maximum value of the Lyapunov exponent satisfies the conjecture (\ref{eq:bound on chaos}) \cite{Hashimoto:2016dfz}. When there is an external force on a particle, the motion of the particle near the horizon of a black hole can be described with the effective Lagrangian. The effective potential of the particle has the extremum point, which is the unstable equilibrium point. The motion of the particle at the extremum point is described as the harmonic oscillator in a non-relativistic limit. 
   The effective Lagrangian of the particle at the extremum point is universally given by the surface gravity $\kappa$ \cite{Hashimoto:2016dfz}:
   \begin{align}
       {\cal L} \sim K(r_0)[\dot{r}^2 + \kappa^2(r - r_0)^2],
       \label{eq:universal lagrangian}
   \end{align}
where $r_0$ is the extremum point of the effective potential and $K(r)$ is a function determined by the metric and test particle. The extremum point is larger than the outer event horizon $r_+$. The equation (\ref{eq:universal lagrangian}) is equivalent to the Lagrangian of harmonic oscillators. Using the equation \ref{eq:universal lagrangian}, the bound on the Lyapunov exponent in a black hole is assumed \cite{Hashimoto:2016dfz},
   \begin{align}
       \lambda \le \kappa\ = \frac{2 \pi T_{BH}}{\hbar}.
       \label{eq:bh conjecture}
   \end{align}
 The surface gravity of the KNdS black hole at the event horizon and the cosmological horizon are given in \cite{Hawking:1977ya},
    \begin{align}
		\kappa = \frac{\Lambda(r_+ - r_{--})(r_+ - r_{-})(r_{c} - r_{+})}{6\Xi({r_+}^2 + a^2)},
    \label{eq:surface gravity}
    \end{align}
    \begin{align}
		\kappa_c = \frac{\Lambda(r_c - r_{--})(r_c - r_{-})(r_{c} - r_{+})}{6\Xi({r_c}^2 + a^2)}.
    \end{align}
 Now, we compare the surface gravity $\kappa$ and Lyapunov exponent $\lambda$ in the KNdS black hole.
In our paper, we use the squared form $\lambda^2\le \kappa^2$ to determine whether the bound (\ref{eq:bh conjecture}) is satisfied or violated.

    \section{Effective Lagrangian and Lyapunov exponent in KNdS black holes}
    \noindent
    In this section, we calculate the maximum Lyapunov exponent of test particles in the KNdS black hole. In addition, we analytically calculate the bound, $\lambda^2\le \kappa^2$, with conditions, the low spin-charge limit, near-horizon limit, and near cosmological horizon limit.

\subsection{Lyapunov exponent of a charged particle}
\noindent
   We start with the Polyakov-type action of a particle on curved space. We assume the particle has an electric charge and angular momentum:
   \begin{align}
       S=\int \left[\frac{1}{2e(X(s))}g_{\mu\nu}(X(s))\dot{X}^\mu(s)\dot{X}^\nu(s)-\frac{e(X(s))}{2}m^2+qA_\mu(X(s))\dot{X}(s)^\mu \right]  ds,
   \end{align}
   where $m$ and $q$ are respectively the mass and charge of the particle, and $e$ is the auxiliary field. $s$ is a parameter that characterizes the geodesic of the particle. In this work, we choose the static gauge, $X^0=s$. Then, the action in the KNdS black hole (\ref{eq:KN_metric_static}) becomes 
        \begin{align}
		S &=\int \Bigg[\frac{1}{2e}\Bigg[-\frac{\Delta_\theta}{\rho^2\Xi^2}\left\{\Delta_r\Delta_\theta-a^2\left(1-\frac{r^2}{3}\Lambda\right)^2\sin^2\theta\right\}+\frac{2a\Delta_\theta\sin^2\theta}{\rho^2\Xi^2}\left\{\Delta_r-(r^2+a^2)\left(1-\frac{r^2}{3}\Lambda\right)\right\}\dot{\Phi} \nonumber \\
		&+\frac{\sin^2\theta}{\rho^2\Xi^2}\left\{\Delta_\theta(r^2+a^2)^2-a^2\Delta_r\sin^2\theta\right\}\dot{\Phi}^2+\frac{\rho^2}{\Delta_r}\dot{r}^2+\frac{\rho^2}{\Delta_\theta}\dot{\theta}^2\Bigg] -\frac{e}{2}m^2-q\frac{Qr\Delta_\theta}{\rho^2\Xi}+q\frac{Qar\sin^2\theta}{\rho^2\Xi}\dot{\Phi}\Bigg] ds,
        \end{align}
where the dots refer to the derivative of $s$. For simplicity, we choose the equatorial motion with $\theta=\pi/2$. The equatorial motion makes $\rho^2=r^2$ and $\Delta_\theta=1$. Then, the Lagrangian of the equatorial motion is
    \begin{align}
		{\cal L}&=\frac{1}{2e}\Bigg[-\frac{1}{r^2\Xi^2}\left\{\Delta_r-a^2\left(1-\frac{r^2}{3}\Lambda\right)^2\right\}+\frac{2a}{r^2\Xi^2}\left\{\Delta_r-(r^2+a^2)\left(1-\frac{r^2}{3}\Lambda\right)\right\}\dot{\Phi} \nonumber \\
		&\qquad+\frac{1}{r^2\Xi^2}\left\{(r^2+a^2)^2-a^2\Delta_r\right\}\dot{\Phi}^2+\frac{r^2}{\Delta_r}\dot{r}^2\Bigg]-\frac{e}{2}m^2-q\frac{Q}{r\Xi}+q\frac{Qa}{r\Xi}\dot{\Phi}.
    \label{eq:equatorial Lagrangian}
    \end{align}
We define an angular momentum in the $\Phi$-direction to obtain the effective Lagrangian using the angular momentum.
    \begin{align}
		L= \frac{\partial {\cal L}}{\partial \dot{\Phi}} =\frac{1}{e}\Bigg[\frac{a}{r^2\Xi^2}\left\{\Delta_r-(r^2+a^2)\left(1-\frac{r^2}{3}\Lambda\right)\right\}+\frac{1}{r^2\Xi^2}\left\{(r^2+a^2)^2-a^2\Delta_r\right\}\dot{\Phi}\Bigg]+q\frac{Qa}{r\Xi}.
    \end{align}
The Lagrangian (\ref{eq:equatorial Lagrangian}) is invariant under the transformation $\Phi\to\Phi+\alpha$. Furthermore, we can reduce the auxiliary field $e$ in the Lagrangian. Using the equation of motion for $e$, $\partial{\cal L}/\partial{e} = 0$, we have
     \begin{align}
		e^2&=-\frac{1}{m^2}\Bigg[-\frac{1}{r^2\Xi^2}\left\{\Delta_r-a^2\left(1-\frac{r^2}{3}\Lambda\right)^2\right\}+\frac{2a}{r^2\Xi^2}\left\{\Delta_r-(r^2+a^2)\left(1-\frac{r^2}{3}\Lambda\right)\right\}\dot{\Phi} \nonumber \\
		&\qquad+\frac{1}{r^2\Xi^2}\left\{(r^2+a^2)^2-a^2\Delta_r\right\}\dot{\Phi}^2+\frac{r^2}{\Delta_r}\dot{r}^2\Bigg].
    \end{align}
 The effective Lagrangian is defined as
	\begin{align}
		{\cal L}_{\rm eff}=&{\cal L}-L\dot{\Phi}.
        \label{eq: effective lagrangian}
	\end{align}
  We assume that the initial velocity of the test particle is small and not relativistic for the radial direction, $\dot{r}^2\ll1$. Then, we can describe the (\ref{eq: effective lagrangian}) as
	\begin{align}
		{\cal L}_{\rm eff}=\frac{1}{2}K(r)\dot{r}^2-V_{\rm eff}(r)+O(\dot{r}^4),
	\label{eq:eff_Lagrangian}
	\end{align}
where
	\begin{align}
		K(r)=\frac{r}{\Delta_r^{3/2}}\left\{ (Lr\Xi - aQq)^2 + m^2(a^4 + r^4)  + a^2m(2Mr - Q^2 - a^2 + r^2)\right\}^{1/2}
	\end{align}
and 
	\begin{align}
		V_{\rm eff}(r)= \frac{La\left\{\Delta_r + (a^2+r^2)(\frac{r^2}{3}\Lambda-1) \right\}- Qqr(a^2 + r^2) - \Delta_r^{2}K(r)}
{a^2\Delta_r - (a^2 + r^2)^2}.
	\label{eq:eff_potential}
	\end{align}
 
 $V_{\rm eff}$ is the effective potential for the test particle in the KNdS black hole. We reduce the constant terms in the equation (\ref{eq:eff_potential}). We are interested in the motion of the test particle at the extremum point $r_0$, where $V'_{\rm eff}(r_0) = 0$. The prime describes the derivative of the radial coordinate $r$. We expand the effective potential $V_{\rm eff}$ at $r_0$.
        \begin{align}
            V_{\rm eff}(r) = V_{\rm eff}(r_0) + \frac{1}{2}V''_{\rm eff}(r_0) (r - r_0)^2
        \end{align}
        
We focus on the region near the extremum point, so we introduce a parameter $\epsilon$ for a small perturbation, $r(s)=r_0+\epsilon(s)$. Then, the effective Lagrangian near the extremum point $r_0$ is described as
    \begin{align}
            {\cal L}_{\rm eff} = \frac{1}{2} K(r_0) \left( \dot{\epsilon}^2 - \frac{V''_{\rm eff}(r_0)}{K(r_0)} \epsilon^2 \right).
    \end{align}
The Lagrangian is also described in a different form according to the equation (\ref{eq: effective lagrangian}):
 	\begin{align}
		{\cal L}_{\rm eff}=\frac{1}{2}K(r_0)\left(\dot{\epsilon^2}+\lambda^2\epsilon^2\right).
	\end{align}
Now, we find the maximum Lyapunov exponent in the KNdS black hole.
 	\begin{align}
		\lambda^2=\left.-\frac{V''_{\rm eff}(r)}{K(r)}\right|_{r=r_0}
        \label{eq:KNdS exponent}
	\end{align}
 The equation (\ref{eq:KNdS exponent}) is equivalent to what was previously reported \cite{Gwak:2022cha} other than the cosmological constant $\Lambda$ is positive. This is expected at first because the metrics for AdS and dS spacetime are the same as the sign of the cosmological constant. However, the positive cosmological constant should make the properties of a test particle different from AdS spacetime.

\subsection{Algebraic analysis}

Seven variables are used to determine the Lyapunov exponent of a test particle, $a, M, Q, q, L, m$, and $\Lambda$. We cannot generally solve the equation $V'_{\rm eff}(r_0) = 0$ to obtain the extremum point $r_0$. However, with specific conditions, we can solve the equation $V'_{\rm eff}(r_0) = 0$ analytically. First, we calculate the Lyapunov exponent of a massless test particle with $Q \to 0$, $a\to 0$, $m = 0$, and $\Lambda < 1/9M^2$. We refer to this as the low spin-charge limit. With the low spin-charge limit, the effective potential of a test particle becomes
    \begin{align}
        V_{\rm eff}' = \frac{\sqrt{3L^2}(3M - r)}{r^2\sqrt{r(-r^3 \Lambda + 3r - 6M)}}.
        \label{eq:V_eff sch}
    \end{align}
 The extremum point $r_0$ is located at $r=3M$. We note that the extremum point $r_0$ is only determined by the mass parameter of the black hole. Furthermore, with zero angular momentum, the extremum point $r_0$ does not exist, which means that chaos does not exist. Therefore, we can say that the angular momentum of a test particle makes the motion chaotic in the low spin-charge limit.
 
 The Lyapunov exponent of the test particle at the extremum point $r_0$ is
    \begin{align}
        \lambda^2 = \frac{1}{27M^2} - \frac{\Lambda}{3}.
        \label{eq:Lyapunov for sch}
    \end{align}
There are no variables that affect the Lyapunov exponent except the mass parameter of the black hole and cosmological constant. When $\Lambda \to 1/9M^2$, the Lyapunov exponent asymptotically approaches zero, but when $\Lambda < 1/9M^2$, it is always positive. Furthermore, massless particles have the same Lyapunov exponent although they have different values of angular momentum. The surface gravity at the event horizon with the low spin-charge limit is
    \begin{align}
        \kappa = \frac{1-\Lambda r_+^2}{2r_+}.
    \end{align}
    The event horizon $r_+$ is obtained by solving $3r -2r_0 - \Lambda r^3 =0$ \cite{Hawking:1998kw}. We substitute $r_0 = 3M$ to compare the surface gravity and the Lyapunov exponent. We can rewrite the surface gravity and the cosmological constant as
    \begin{align}
        \kappa =  \frac{1}{r_+} \left(\frac{r_0}{r_+} - 1 \right), \quad \Lambda=\frac{3r_+ - 2r_0}{r_+^3}
    \end{align}
Because the cosmological constant is positive, we can consider that $0 < 3r_+-2r_0$. The bound is calculated as
    \begin{align}
        \kappa^2 - \lambda^2 &= \frac{1}{r_+^2} \left(\frac{r_0^2}{r_+^2} - \frac{2r_0}{r_+} +1 \right) - \left(\frac{1}{3r_0^2} - \frac{1}{r_+^2} + \frac{2r_0}{3r_+^3} \right) \nonumber \\
        &=\frac{1}{{r_0}^2} \left( \frac{{r_0}^4}{{r_+}^4} - \frac{8{r_0}^3}{3{r_+}^3} + \frac{2{r_0}^2}{{r_+}^2} 
        -\frac{1}{3}\right) >0.
    \label{eq: sch bound}
    \end{align}
 The condition of $r_0$ is $r_+ < r_0 < \frac{3}{2}r_+$, and we want to know whether the bound in the low spin-charge limit is positive or negative. By deriving the bound (\ref{eq: sch bound}) with $r_0$, we find that the equation (\ref{eq: sch bound}) increases when $r_0 > r_+$. When $r_0 \to r_+$, the minimum value is zero. Thus, the bound (\ref{eq: sch bound}) is always satisfied in the range $r_+ < r_0 < \frac{3}{2}r_+$ with the low spin-charge limit.
    
    Second, we calculate the near-horizon limit, $r_0 \to r_+$. For this limit, we assume that the extremum point of the effective potential appears near the event horizon $r_+$. With the near-horizon limit, we do not solve the equation $V'_{\rm eff}(r_0) = 0$. However, we directly set the limit $r_0 \to r_+$ in the equation (\ref{eq:KNdS exponent}). With the near-horizon limit, the Lyapunov exponent is rewritten as
    \begin{align}
        \lambda^2 |_{r_0 \to r_+} = \Lambda^2\frac{(r_c - r_+)^2(r_{--} - r_+)^2(r_- - r_+)^2}{36(a^2 + {r_+}^2)^2} = \kappa^2 \Xi^2,
        \label{eq:near horizon exponent}
    \end{align}
    and the bound is
    \begin{align}
        \kappa^2 - \lambda^2 |_{r_0 \to r_+} = \kappa^2(1-\Xi^2).
        \label{eq:near horizon bound}
    \end{align}
Considering the equation (\ref{eq:Xi}), the bound with the near-horizon limit is violated when the cosmological constant is positive. However, the bound with the near-horizon limit is satisfied when the cosmological constant is negative. The result with the near-horizon limit matches the result of the prior research \cite{Gwak:2022cha}. In addition, we find a similar result with the near cosmological horizon limit, $r_0 \to r_c$. In this limit, we assume the extremum point of the effective potential $r_0$ is near the cosmological horizon $r_c$. The Lyapunov exponent (\ref{eq:KNdS exponent}) includes $\Delta_r$. Considering that $\Delta_r = 0$ at $r = r_+$ and $r = r_c$, we find the near cosmological horizon limit from the equation (\ref{eq:near horizon exponent}). With the near cosmological horizon limit, the Lyapunov exponent is
    \begin{align}
        \lambda^2 |_{r_0 \to r_c} = \Lambda^2\frac{(r_c - r_+)^2(r_{c} - r_-)^2(r_{c} - r_{--})^2}{36(a^2 + {r_c}^2)^2} = {\kappa_c}^2 \Xi^2.
    \end{align}
The bound with the near cosmological horizon limit is
    \begin{align}
        \kappa^2 - \lambda^2 |_{r_0 \to r_c} &= \kappa^2-{\kappa_c}^2 \Xi^2. \nonumber \\
        &= \Lambda^2\frac{(r_c - r_+)^2}{36} \left[ \frac{(r_+ - r_{--})^2(r_+ - r_-)^2}{\Xi^2 (a^2 + {r_+}^2)^2} - \frac{(r_{c} - r_-)^2(r_c - r_{--})^2}{(a^2 + {r_c}^2)^2} \right]
        \label{eq:near cosmological horizon bound}
    \end{align}
    The bound with the near cosmological horizon limit exists when the cosmological constant is positive. Whether the bound (\ref{eq:near cosmological horizon bound}) is satisfied or not is determined by the parameters of the black hole. Thus, although the distance between $r_0$ and $r_+$ is so large, we cannot determine whether the bound is satisfied or not. According to Equations (\ref{eq:near horizon bound}) and (\ref{eq:near cosmological horizon bound}), the bound is not linearly dependent on the distance between $r_+$ and $r_0$ in KNdS black hole. Therefore, in order to probe the Lyapunov exponent in the KNdS black hole, we need to use the numerical method.

    \section{Bound for RNdS black holes}

 We numerically analyze the maximum Lyapunov exponent in the RNdS black hole. The Lyapunov exponent for a test particle in the RNdS black hole is
    \begin{align}
        \lambda^2 = -\frac{2Qq{\Delta_r}^{3/2}}{{r_0}^{4}\sqrt{L^{2} {r_0}^{2} + m^{2} {r_0}^4}}+\frac{\Delta_{r}}{{r_0}^{5}}(4\Delta'_{r} + 2\Lambda {r_0}^{3} - {r_0})- \frac{\Delta_r^{2}}{{r_0}^{6}}(6 + \frac{5m^{2} {r_0}^{4}}{L^{2} {r_0}^{2} + m^{2} {r_0}^{4}})+ \frac{{\Delta'}_r^{2}}{2{r_0}^{4}},
    \label{eq:RN exponent}
    \end{align}
 which is equal to Equation (\ref{eq:KNdS exponent}) with zero spin parameter, $a \to 0$.
 \begin{figure}[hp]
    \centering
    \begin{subfigure}{0.45\textwidth}
        \centering
        \includegraphics[width=1.1\linewidth]{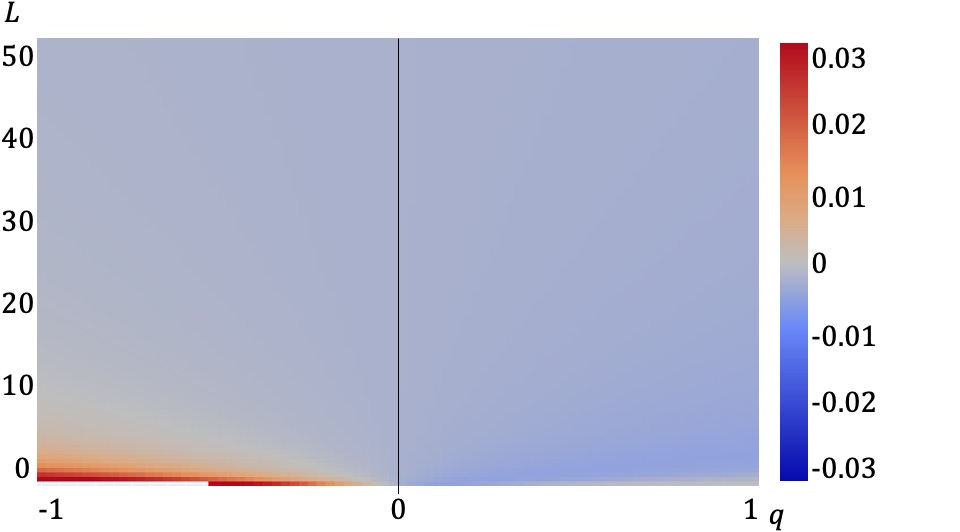}
        \caption{($Q,\Lambda$) = $(0.95,0)$}
        \label{fig:graph_a}
    \end{subfigure}
    \hfill
    \begin{subfigure}{0.45\textwidth}
        \centering
        \includegraphics[width=1.1\linewidth]{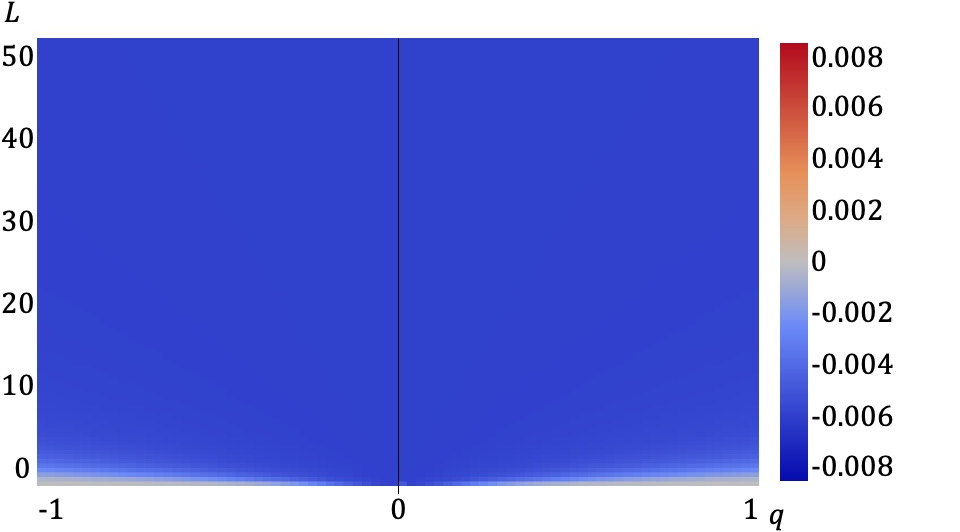}
        \caption{($Q,\Lambda$) = $(1,0.1)$}
        \label{fig:graph_b}
    \end{subfigure}
    
    \vskip\baselineskip
    
    \begin{subfigure}{0.45\textwidth}
        \centering
        \includegraphics[width=1.1\linewidth]{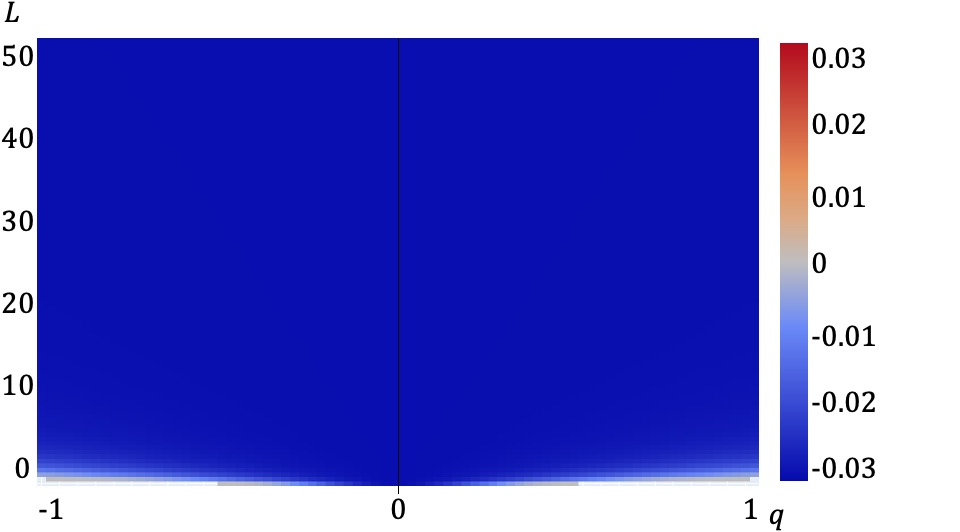}
        \caption{Extremal ($Q,\Lambda$) = $(1,0)$}
        \label{fig:graph_c}
    \end{subfigure}
    \hfill
    \begin{subfigure}{0.45\textwidth}
        \centering
        \includegraphics[width=1.1\linewidth]{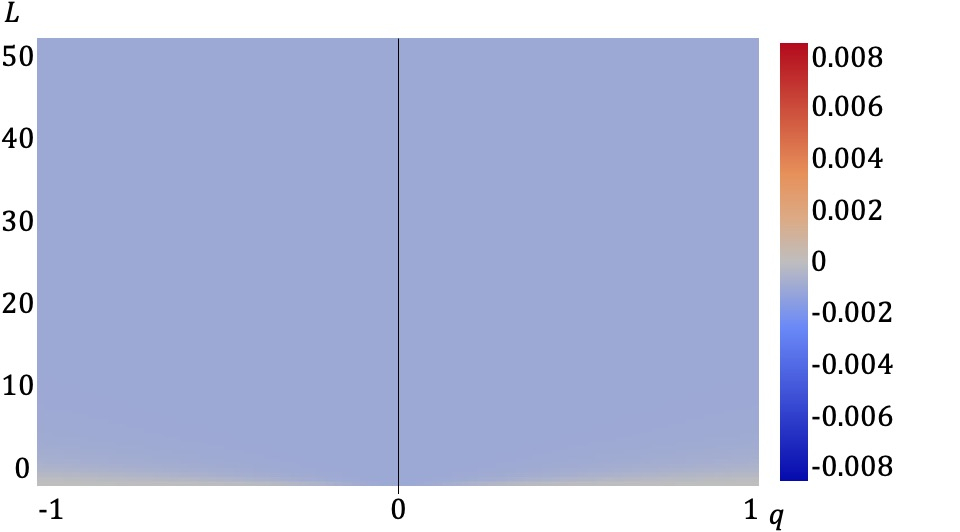}
        \caption{($Q,\Lambda$) = $(1,0.15)$}
        \label{fig:graph_d}
    \end{subfigure}
    \vskip\baselineskip
    \begin{subfigure}{0.45\textwidth}
        \centering
        \includegraphics[width=1.1\linewidth]{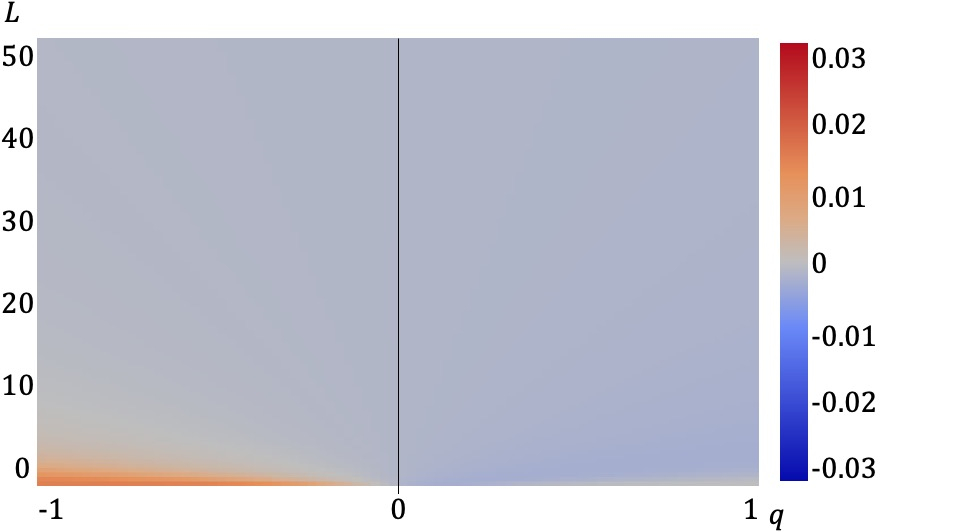}
        \caption{($Q,\Lambda$) = $(0.96,0.05)$}
        \label{fig:graph_e}
    \end{subfigure}
    \hfill
    \begin{subfigure}{0.45\textwidth}
        \centering
        \includegraphics[width=1.1\linewidth]{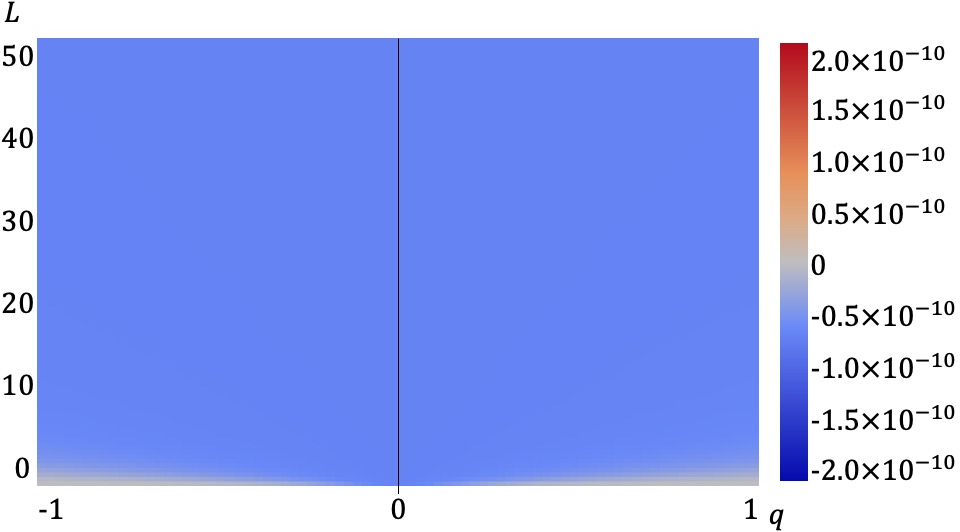}
        \caption{near Nariai ($Q,\Lambda$) = $(1,0.19)$}
    \end{subfigure}
    \vskip\baselineskip
    \begin{subfigure}{0.45\textwidth}
        \centering
        \includegraphics[width=1.1\linewidth]{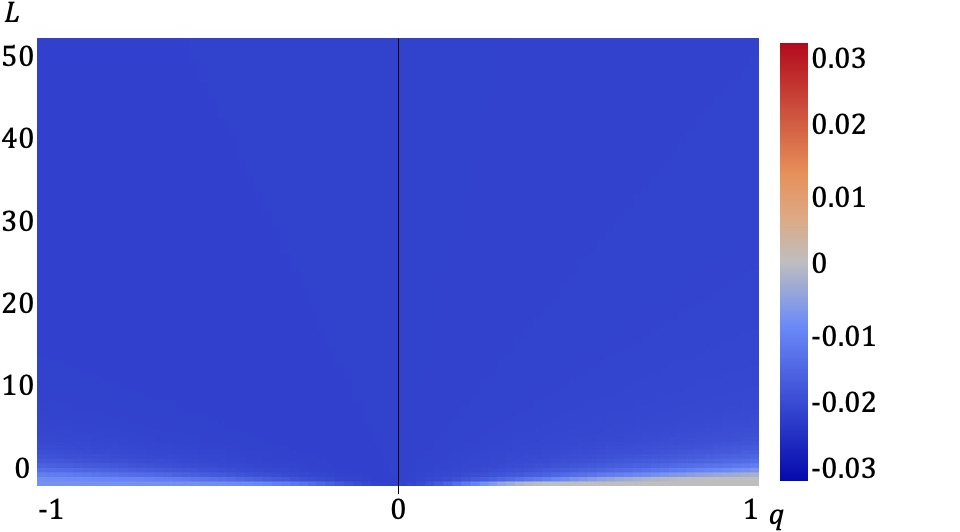}
        \caption{Extremal ($Q,\Lambda$) = $(1.01,0.05)$}
        \label{fig:graph_g}
    \end{subfigure}
    \hfill
    \begin{subfigure}{0.45\textwidth}
        \centering
        \includegraphics[width=1.1\linewidth]{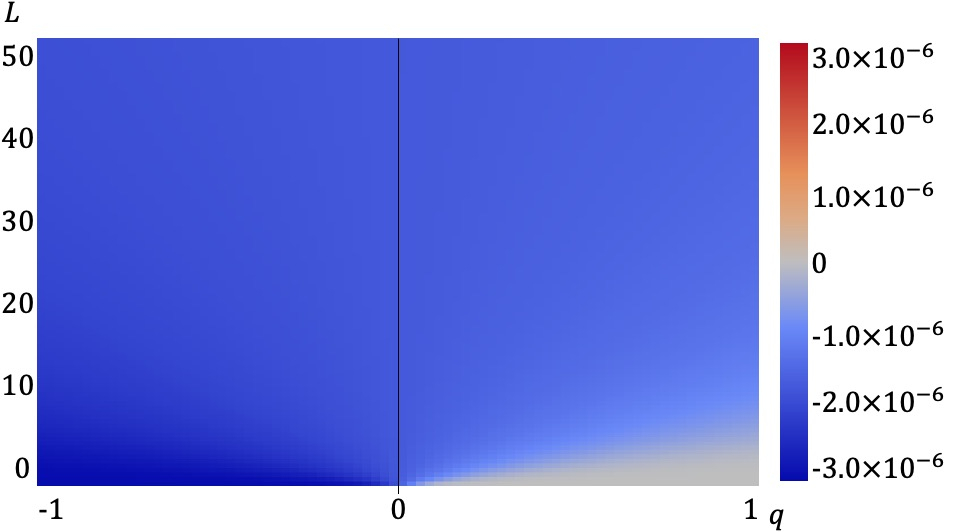}
        \caption{near Ultracold ($Q,\Lambda$) = $(1.06,0.22)$}
    \end{subfigure}
    \caption{The result for the particle with $m=0$ in the RNdS black hole.}
    \label{fig:m=0_RN}
\end{figure}

\begin{figure}[hp]
    \centering
    \begin{subfigure}{0.45\textwidth}
        \centering 
        \includegraphics[width=1.1\linewidth]{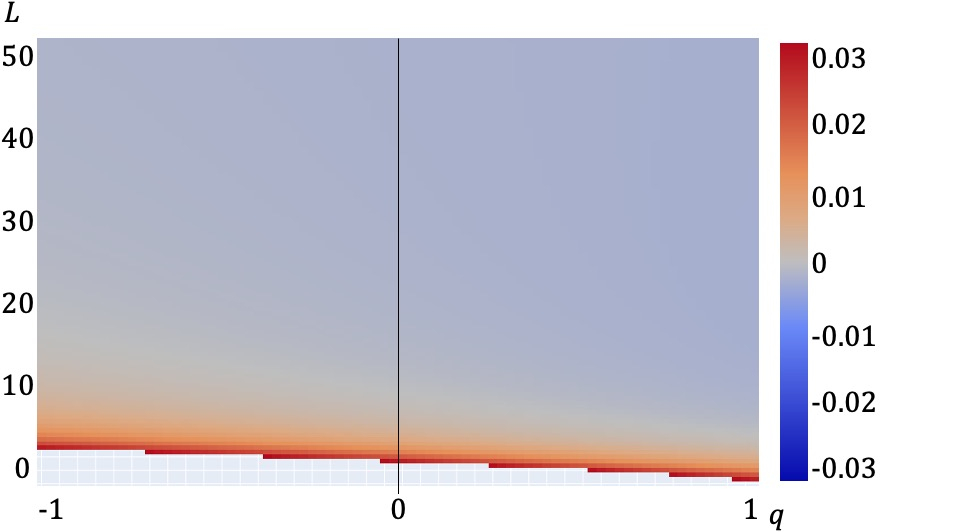}
        \caption{($Q,\Lambda$) = $(0.95,0)$}
    \end{subfigure}
    \hfill
    \begin{subfigure}{0.45\textwidth}
        \centering
        \includegraphics[width=1.1\linewidth]{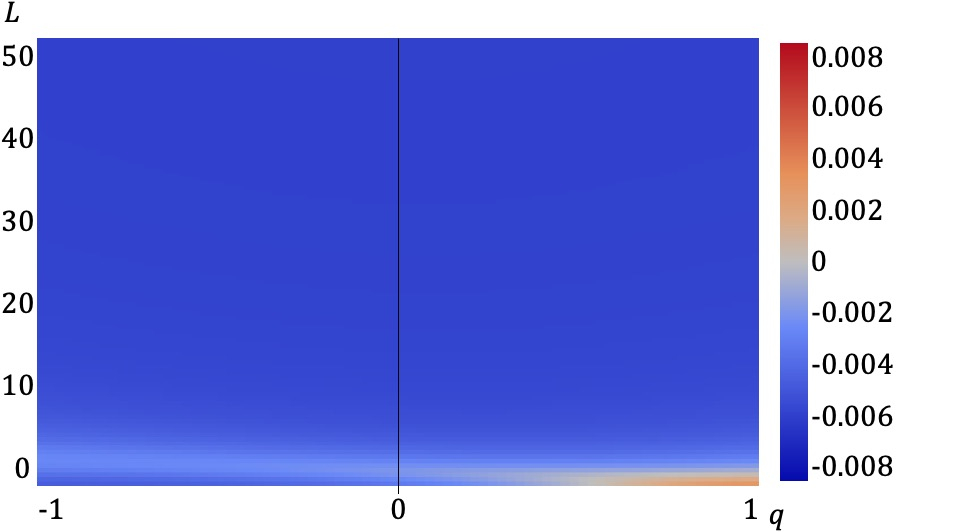}
        \caption{($Q,\Lambda$) = $(1,0.1)$}
    \end{subfigure}
    \vskip\baselineskip
    \begin{subfigure}{0.45\textwidth}
        \centering
        \includegraphics[width=1.1\linewidth]{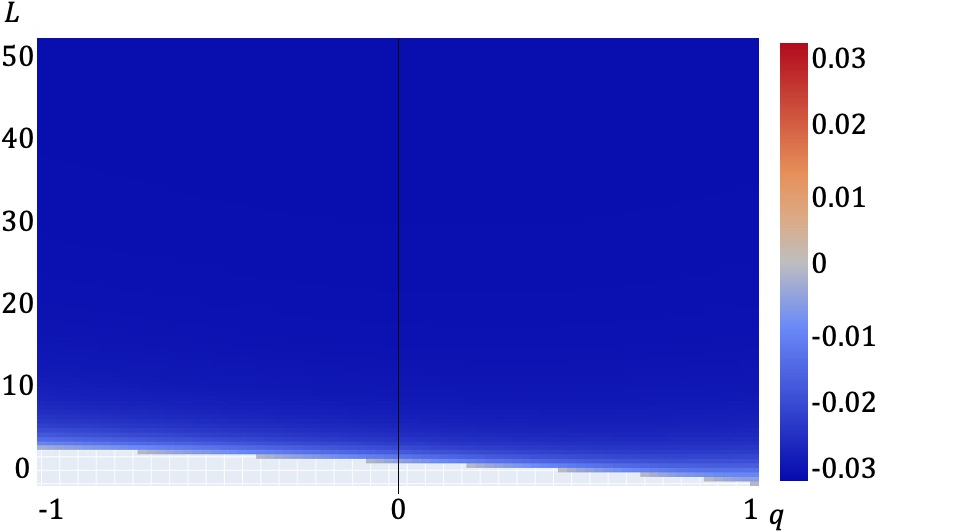}
        \caption{Extremal ($Q,\Lambda$) = $(1,0)$}
    \end{subfigure}
    \hfill
    \begin{subfigure}{0.45\textwidth}
        \centering
        \includegraphics[width=1.1\linewidth]{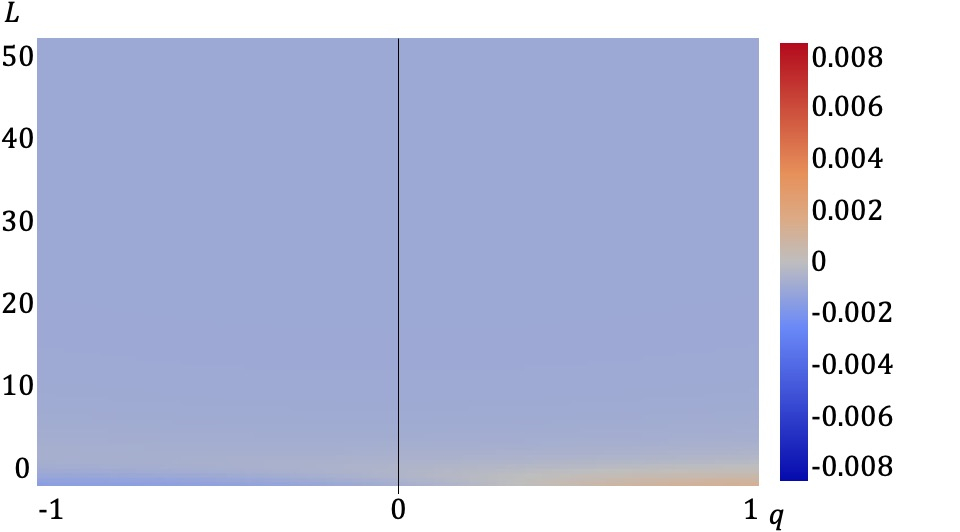}
        \caption{($Q,\Lambda$) = $(1,0.15)$}
    \end{subfigure}
    \vskip\baselineskip
    \begin{subfigure}{0.45\textwidth}
        \centering
        \includegraphics[width=1.1\linewidth]{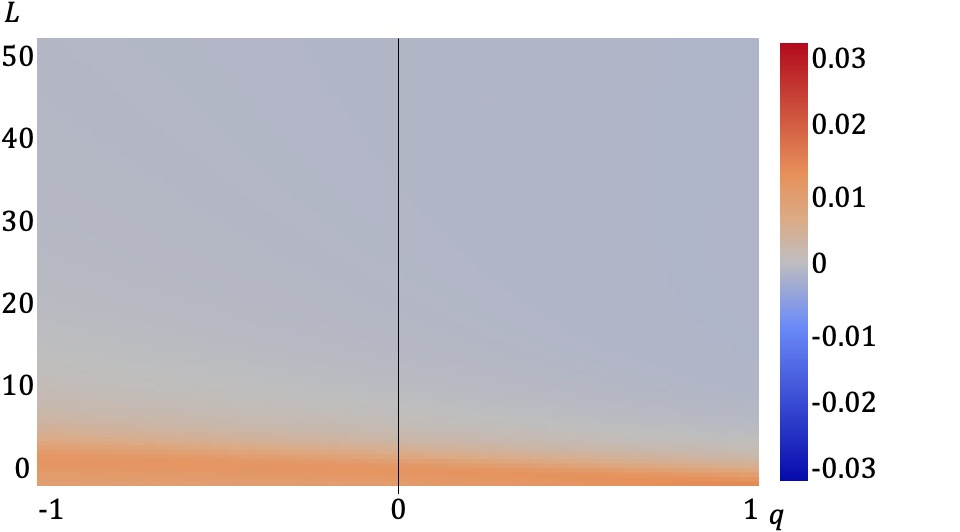}
        \caption{($Q,\Lambda$) = $(0.96,0.05)$}
    \end{subfigure}
    \hfill
    \begin{subfigure}{0.45\textwidth}
        \centering
        \includegraphics[width=1.1\linewidth]{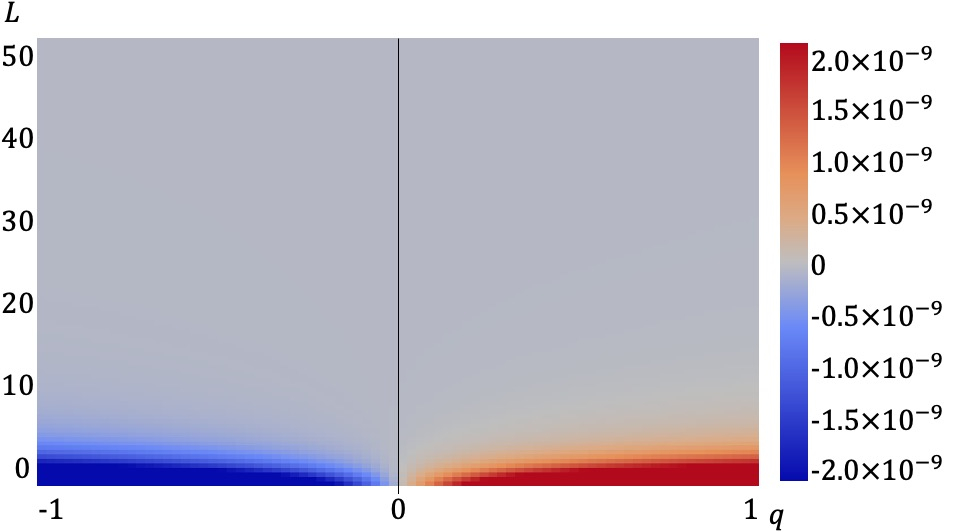}
        \caption{near Nariai ($Q,\Lambda$) = $(1,0.19)$}
    \end{subfigure}
    \vskip\baselineskip
    \begin{subfigure}{0.45\textwidth}
        \centering
        \includegraphics[width=1.1\linewidth]{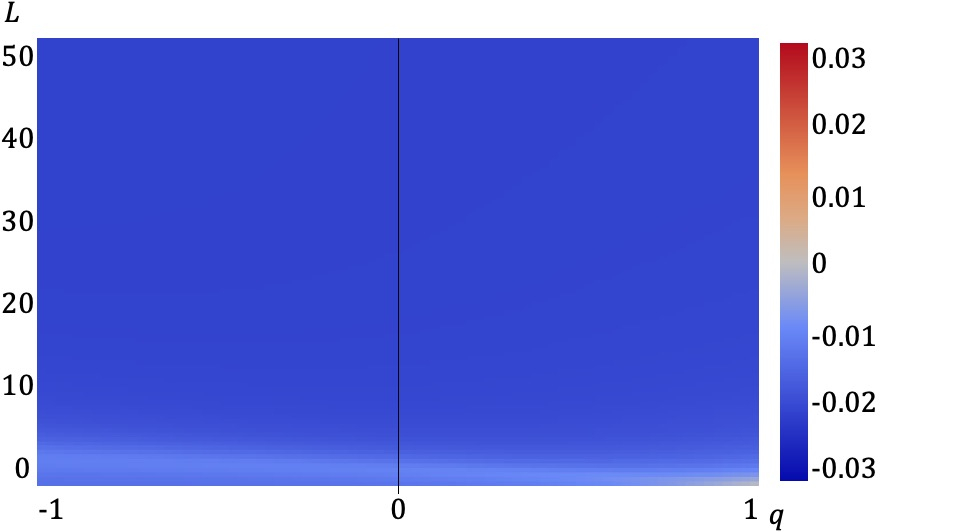}
        \caption{Extremal ($Q,\Lambda$) = $(1.01,0.05)$}
    \end{subfigure}
    \hfill
    \begin{subfigure}{0.45\textwidth}
        \centering
        \includegraphics[width=1.1\linewidth]{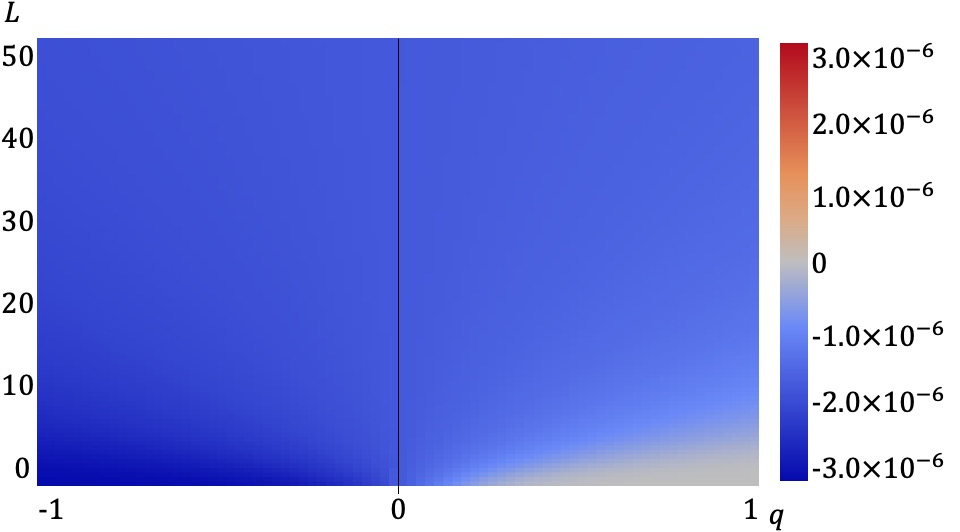}
        \caption{near Ultracold ($Q,\Lambda$) = $(1.06,0.22)$}
    \end{subfigure}
    \caption{The result for the particle with $m=1$ in the RNdS black hole.}
    \label{fig:m=1_RN}
\end{figure}

  We plot eight graphs for the bound, $\kappa^2 - \lambda^2$, of massless test particles in Fig.\ref{fig:m=0_RN}. We use the dimensionless parameter, $M = 1$, in the numerical calculation. The horizontal and vertical axes are respectively the electric charge and angular momentum of a test particle. We only consider the positive angular momentum because the angular momentum appears in a square form in Equation (\ref{eq:RN exponent}). Considering that the RNdS black hole is static, it is reasonable that the direction of the angular momentum does not affect the Lyapunov exponent. We classify the regions by color depending on whether the bound is satisfied or not. If the bound is satisfied at a point, $\kappa^2 - \lambda^2 \ge 0$, we plot the point in red. Conversely, if the bound is violated, $\kappa^2 - \lambda^2 < 0$, the point is described in blue. The magnitude of the bound, $|\kappa^2 - \lambda^2|$, is described according to the brightness of color. The point with the smaller absolute value of the bound is described as being lighter than the other point. When the bound is near zero, the point looks gray. We can check the scale of each graph with the color bar on the right side. We plot the point in white where the Lyapunov exponent is negative. This means that the chaos does not exist in the white region. We denote the parameter of the black hole $(Q, \Lambda)$ below the graph. In addition, if the graph is for the extremal case, we denote it as well. For the Nariai and ultracold black holes, we calculate the near Nariai limit and near ultracold limit. Fig.\ref{fig:m=1_RN} shows plots of graphs for massive test particles with $m =1$.

  First, we explain Fig.\ref{fig:m=0_RN}. We divide the graphs into two columns. On the left column, we display the RN black hole with $\Lambda = 0$ and the RNdS black hole with $\Lambda = 1/20$ to focus on the effect of the positive cosmological constant.
  Fig.\ref{fig:m=0_RN} (a) is the graph for the flat RN black hole with $Q = 0.95$. With a low angular momentum, it shows a blue region when the charge of the test particle is positive and a red when the charge is negative. Further, there is a white region when the charge is negative. When the angular momentum increases, the color becomes light and all the points violate the bound, $\kappa^2 - \lambda^2$.
  Fig.\ref{fig:m=0_RN} (c) is the graph for the extremal flat RN black hole with $Q = 1$. There is no red region in Fig.\ref{fig:m=0_RN} (c). This is because the surface gravity of the extremal black hole is zero. With a low angular momentum and negative charge of a test particle, it shows the white region. When the angular momentum increases, the chaos appears and the Lyapunov exponent also increases.
  Fig.\ref{fig:m=0_RN} (e) is the graph for the RNdS black hole with $Q = 0.96$ and $\Lambda = 0.05$. It appears to be similar to Fig.\ref{fig:m=0_RN} (a), but the white region with a low angular momentum disappears. When the angular momentum increases, the bound becomes violated with a negative charge of a test particle.
  Fig.\ref{fig:m=0_RN} (g) is the graph for the extremal RNdS black hole with $Q = 1.01$ and $\Lambda = 0.05$. This appears to be similar to Fig.\ref{fig:m=0_RN} (c), but the white region with a low angular momentum disappears. The Lyapunov exponent increases as the angular momentum increases, so the color becomes more vivid.
  In the left column, we find that the positive Lyapunov exponent makes the white region disappear, which means that the motion of a test particle becomes more chaotic. The large angular momentum makes the bound violated, but it does not guarantee that the Lyapunov exponent increases when the angular momentum increases. By focusing on the region with the positive charge in Fig.\ref{fig:m=0_RN} (a) and (e), we determine that the Lyapunov exponent decreases as the angular momentum increases.

  On the right column, we set the charge parameter, $Q = 1$, varying the cosmological constant until the RNdS black hole becomes the near Nariai limit.  The right column also includes a graph for the near ultracold limit, Fig.\ref{fig:m=0_RN} (h). It should be noted that the scale of the color bars becomes different.
  Fig.\ref{fig:m=0_RN} (b) and (d) are graphs of the RNdS black holes with $\Lambda = 0.1$ and $\Lambda = 0.15$, respectively. Fig.\ref{fig:m=0_RN} (f) is the graph for the near Nariai limit with $\Lambda =0.18749$. It becomes the Nariai black hole when $\Lambda =0.1875$. In these graphs, all of the regions are described in blue. The Lyapunov exponent increases as the angular momentum increases. If the absolute values of the charge of test particles are the same, the bound also has the same value in these graphs.
  Fig.\ref{fig:m=0_RN} (h) is the graph for the near ultracold black hole with $Q = 1.06$ and $\Lambda = 199/900$. When $\Lambda = 2/9$, it becomes an ultracold black hole. All the regions in Fig.\ref{fig:m=0_RN} (h) violate the bound. However, when the angular momentum is low, the bound is near zero with a positive charge. The Lyapunov exponent increases as the angular momentum increases when the electric charge is positive. Conversely, when the charge is negative, the Lyapunov exponent decreases.

    Now, we explain Fig.\ref{fig:m=1_RN}. The arrangement is the same as Fig.\ref{fig:m=0_RN}. Therefore, if the position is the same as Fig.\ref{fig:m=0_RN}, the electric charge and cosmological constant of the black hole are the same.
  Fig.\ref{fig:m=1_RN} (a) is the graph for the flat RN black hole with $Q = 0.95$. It has a white region with a low angular momentum. The white region becomes narrow as the charge of a test particle increases. In addition, there is a narrow red region where the bound is satisfied with a low angular momentum. With a large angular momentum, all the regions violate the bound.
  Fig.\ref{fig:m=1_RN} (c) is the graph for the extremal RN black hole with $Q = 1$. It has a white region with a low angular momentum. The Lyapunov exponent increases as the angular momentum increases.
  Fig.\ref{fig:m=1_RN} (e) is the graph for the RNdS black hole with $Q = 0.96$ and $\Lambda = 0.05$. It has a red region with a low angular momentum. When the charge of a test particle increases, this red region becomes narrow. The bound is violated with a large angular momentum.
  Fig.\ref{fig:m=1_RN} (g) is the graph for the extremal RNdS black hole with $Q = 1.01$ and $\Lambda = 0.05$. All the regions are described in blue, which means that the bound is always violated. We find the light blue line with the low angular momentum, which means that the Lyapunov exponent first decreases and then increases as the angular momentum increases.

  Likewise, we set the electric charge $Q=1$, varying the cosmological constant in the right column in Fig.\ref{fig:m=1_RN}. Fig.\ref{fig:m=1_RN} (b) and (d) are the graphs for the RNdS black holes with $\Lambda = 0.1$ and $\Lambda = 0.15$, respectively. They have a light blue line with a low angular momentum. In addition, there is a red region with a low angular momentum and positive charge. However, as the angular momentum increases, all the regions become violated.
  Fig.\ref{fig:m=1_RN} (f) is the graph for the near Nariai RNdS black hole with $\Lambda =0.1875$. It has a red region with a low angular momentum and positive charge. However, all the regions violate the bound when the angular momentum increases.
  Fig.\ref{fig:m=1_RN} (h) is the graph for the near ultracold RNdS black hole with $Q = 1.06$ and $\Lambda = 199/900$. It has the same properties as Fig.\ref{fig:m=0_RN} (h).

   By comparing Fig.\ref{fig:m=0_RN} and Fig.\ref{fig:m=1_RN}, we find that the white and red regions become large when the particle is massive. Therefore, we can deduce that the mass of a test particle makes the motion of the particle less chaotic.
  Considering all graphs in Fig.\ref{fig:m=0_RN} and Fig.\ref{fig:m=1_RN}, there are common results. The color of the graphs becomes light as the cosmological constant increases except for the graphs for the near ultracold, Fig.\ref{fig:m=0_RN} (h) and Fig.\ref{fig:m=1_RN} (h). The near ultracold black hole has a large electric charge, which makes the color vivid. A large cosmological constant makes the surface gravity $\kappa$ small and the cosmological horizon $r_c$ near the outer horizon $r_+$. A test particle explores between two horizons $r_+$ and $r_c$. As the cosmological constant increases, the region in which the particle can move becomes narrow. Therefore, we can consider that there is a tendency that the shorter distance between $r_+$ and $r_c$ would make the Lyapunov exponent smaller. Thus, a large cosmological constant makes two terms in the bound small, which makes the color light.    In addition, the test particle with a positive charge has a smaller bound than the test particle with an opposite charge in Fig.\ref{fig:m=0_RN}. For clarity, we consider the electric force between the test particle and the black hole. We consider the situation where the test particle is rotating around the black hole and falls into the black hole. In this case, if the charge of the test particle is positive and the test particle feels the repulsive electric force, the rotating time and the exploring distance of the test particle become longer. The Lyapunov exponent is dependent on the distance according to the definition (\ref{eq:Lyapunov}). As a result, the Lyapunov exponent becomes large. Likewise, the test particle with an opposite charge feels an attractive force, which takes the opposite result.

   Here, we explain the role of the angular momentum in the RNdS black hole. According to Fig.\ref{fig:m=1_RN} (e), a test particle with a low angular momentum shows a red color, but with a large angular momentum, it has a light blue color. It appears that the angular momentum makes the Lyapunov exponent larger and the motion chaotic. Now, we focus on Fig.\ref{fig:m=1_RN} (f). A test particle with a low angular momentum and opposite charge has a blue color and the color becomes lighter as the angular momentum increases. This means that the Lyapunov exponent decreases and that the motion becomes less chaotic. These properties can be found in other graphs in Fig.\ref{fig:m=0_RN} and Fig.\ref{fig:m=1_RN}. Such results are contrary to each other. Moreover, for Fig.\ref{fig:m=1_RN} (b),(d), and (g), we can find the white line at a low angular momentum. This means that as the angular momentum increases, the Lyapunov exponent first decreases and then increases again. Thus, we conclude that there is a non-linear relation between the Lyapunov exponent and the angular momentum in the RNdS black hole.

\section{Bound for KdS black holes}

 In this section, we numerically calculate the Lyapunov exponent and bound, $\kappa^2 - \lambda^2$, in the KdS black hole. In order to calculate the Lyapunov exponent in the KdS black hole, we take the zero-charge limit of Equation (\ref{eq:KNdS exponent}), $Q \to 0$. Because the electric charge of the black hole disappears, we can ignore the effect of the electric force between a particle and a black hole. Therefore, we plot the graphs for the bound, $\kappa^2 - \lambda^2$, varying the spin parameter.
 
 \begin{figure}[hp]
    \centering
    \begin{subfigure}{0.45\textwidth}
        \centering 
        \includegraphics[width=1.1\linewidth]{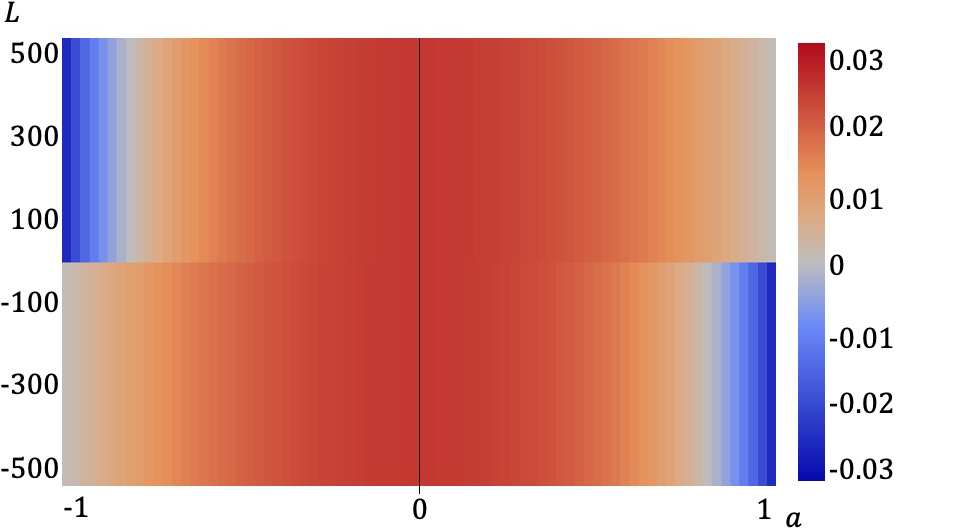}
        \caption{($m,\Lambda$) = $(0,0)$}
    \end{subfigure}
    \hfill
    \begin{subfigure}{0.45\textwidth}
        \centering
        \includegraphics[width=1.1\linewidth]{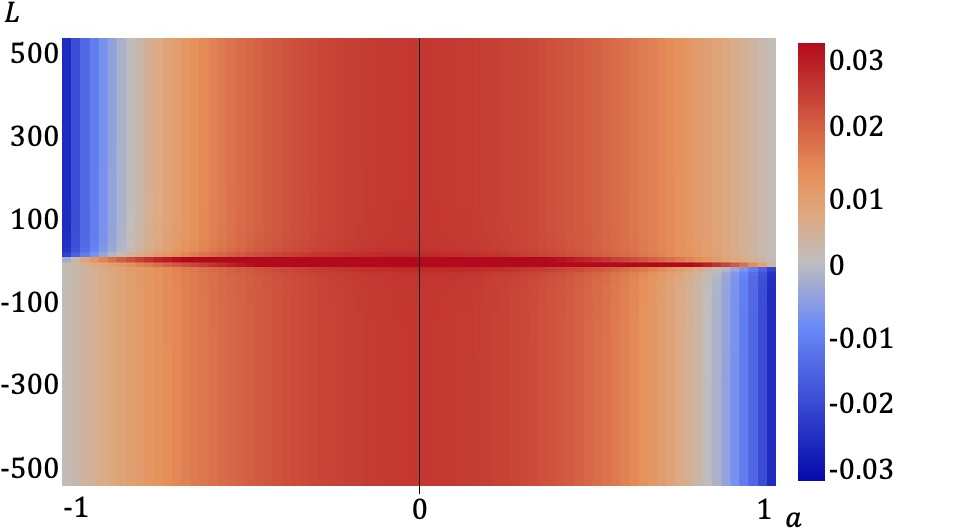}
        \caption{($m,\Lambda$) = $(1,0)$}
    \end{subfigure}
    \vskip\baselineskip
    \begin{subfigure}{0.45\textwidth}
        \centering
        \includegraphics[width=1.1\linewidth]{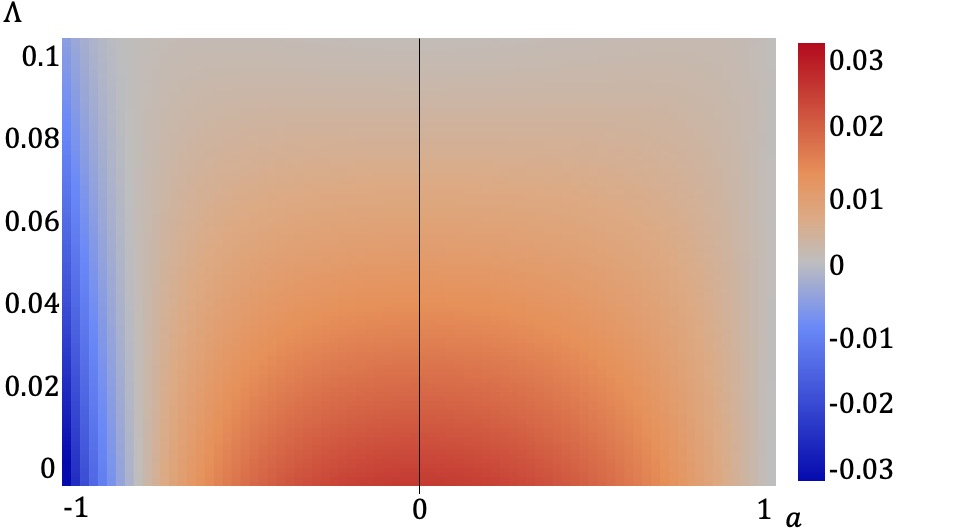}
        \caption{($m$) = $(0)$}
    \end{subfigure}
    \hfill
    \begin{subfigure}{0.45\textwidth}
        \centering
        \includegraphics[width=1.1\linewidth]{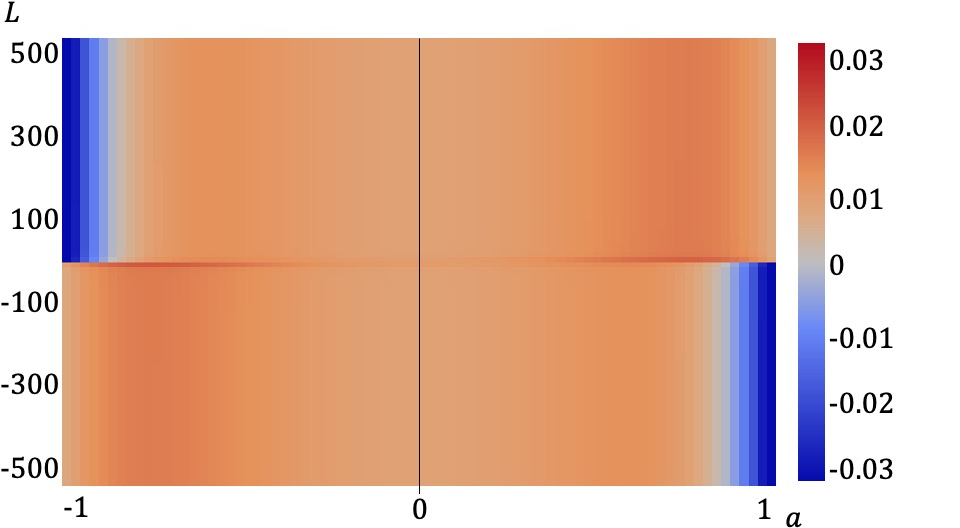}
        \caption{($m,\Lambda$) = $(1,0.1)$}
    \end{subfigure}
    \vskip\baselineskip
    \begin{subfigure}{0.45\textwidth}
        \centering
        \includegraphics[width=1.1\linewidth]{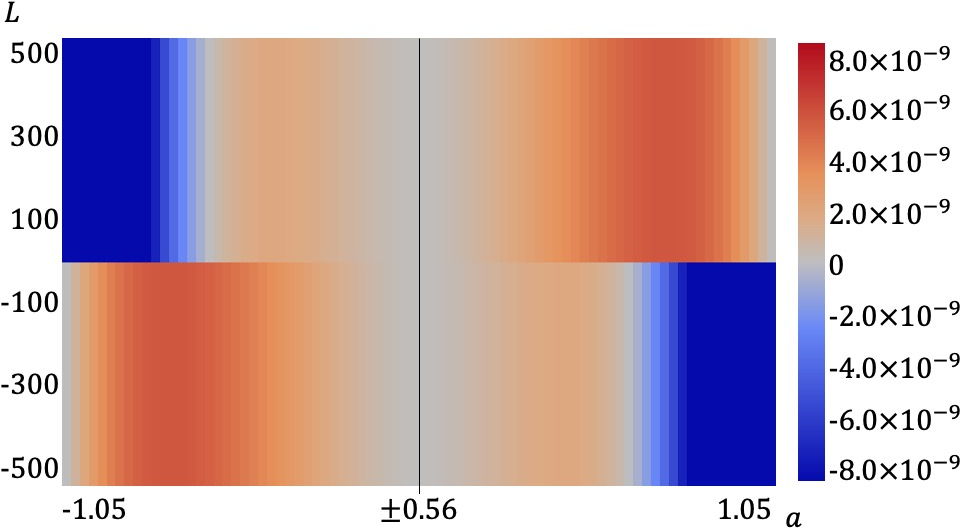}
        \caption{near Nariai ($m,\Lambda$) = $(0,0.12)$}
    \end{subfigure}
    \hfill
    \begin{subfigure}{0.45\textwidth}
        \centering
        \includegraphics[width=1.1\linewidth]{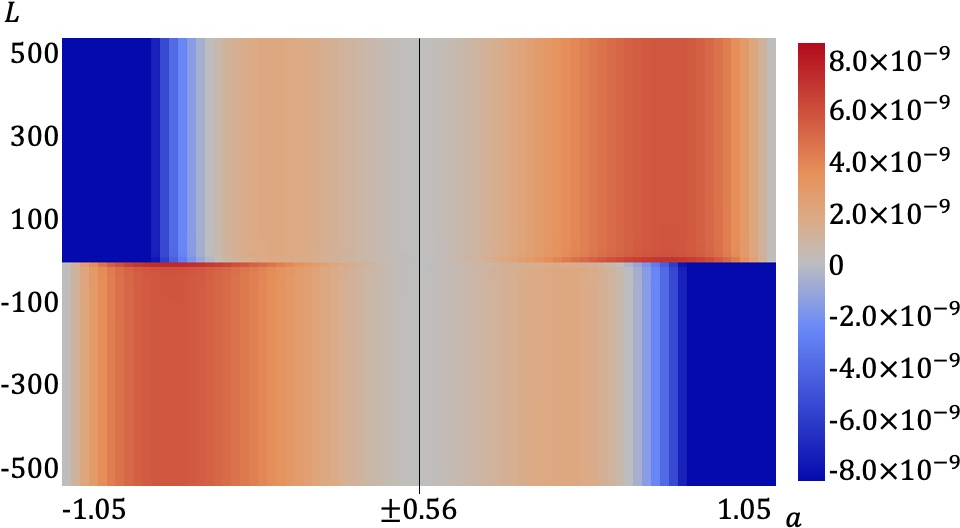}
        \caption{near Nariai ($m,\Lambda$) = $(1,0.12)$}
    \end{subfigure}
    \vskip\baselineskip
    \begin{subfigure}{0.45\textwidth}
        \centering
        \includegraphics[width=0.4\linewidth]{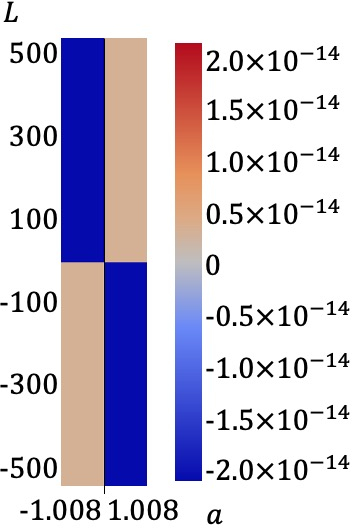}
        \caption{near Ultracold ($m,\Lambda$) = $(0,0.18)$}
    \end{subfigure}
    \hfill
    \begin{subfigure}{0.45\textwidth}
        \centering
        \includegraphics[width=0.4\linewidth]{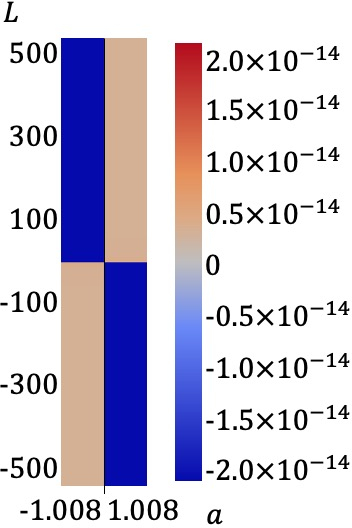}
        \caption{near Ultracold ($m,\Lambda$) = $(1,0.18)$}
    \end{subfigure}
    \caption{The left column is the result for the particle with $m=0$ in the KdS black hole. The right column is the result for the particle with $m=1$.}
    \label{fig:Kerr}
\end{figure}
In Fig.\ref{fig:Kerr}, we plot the eight graphs that show the bound of a test particle in the KdS black hole. We set the mass parameter $M = 1$. The red, blue, and white regions represent the satisfaction, violation of the bound, and the non-chaotic, respectively. We plot the graphs for the bound of a test particle with $m = 0$ in the left column and with $m = 1$ in the right column. The horizontal axis is the spin parameter and the vertical axis is the angular momentum of a test particle except in Fig.\ref{fig:Kerr} (c). In the paragraph below, we explain Fig.\ref{fig:Kerr} (c). Except for Fig.\ref{fig:Kerr} (c), we denote the mass of the test particle and cosmological constant, ($m, \Lambda$), below each graph.

 First, we focus on the graphs in the left column.
 Fig.\ref{fig:Kerr} (a) plots the bound of the massless particle in the flat Kerr black hole, $\Lambda = 0$. The bound is violated only when the directions of the angular momentum of a test particle and the spin parameter of the black hole are opposite to each other. The bound does not vary although the angular momentum becomes different. Thus, Fig.\ref{fig:Kerr} (a) is point-symmetric. We also find these properties in the case of a massless particle in the KdS black hole. In Fig.\ref{fig:Kerr} (c), we therefore set the vertical horizon as the cosmological constant instead of the angular momentum, and denote only the mass of a test particle below Fig.\ref{fig:Kerr} (c). We set the angular momentum to be positive. Regardless of the cosmological constant, the bound is violated only when the spin parameter is large and the direction is opposed to the angular momentum of a test particle. In addition, in Fig.\ref{fig:Kerr} (c), we can find that the large cosmological constant makes the magnitude of the bound, $\kappa^2 - \lambda^2$, decrease and the blue region narrow, which is similar to the RNdS black hole.
 Fig.\ref{fig:Kerr} (e) shows the bound for a massless particle in the KdS black hole with $\Lambda = 0.12$. With $\Lambda = 0.12$, the black hole can exist when $0.56\le a \le 1.05$. Therefore, Fig.\ref{fig:Kerr} (e) includes the near Nariai and extremal black holes. When $a = 0.56$ and $a = 1.05$, the black holes are the near Nariai limit and the extremal black hole, respectively. The bound is violated only when the spin parameter is large and opposed to the angular momentum of a test particle. For the near Nariai black hole, the bound is always satisfied with the near-zero. Although the direction of the spin parameter for the near Nariai is opposite to the angular momentum, the bound is satisfied. The blue region in Fig.\ref{fig:Kerr} (e) appears broader than the blue region in Fig.\ref{fig:Kerr} (a) and (c). However, this is because the range of the spin parameter is different. It should be noted that the scale of the color bar may become different. Considering these, we can find that the large cosmological horizon makes the blue region and the magnitude of the bound smaller.
 Fig.\ref{fig:Kerr} (g) plots the bound in the near ultracold black hole with $\Lambda = 0.18$. According to Fig.\ref{fig:blackhole_region}, when the cosmological constant $\Lambda = 16/9$, the KdS black hole becomes the ultracold black hole on the point. Therefore, we do not vary the spin parameter of the near ultracold KdS black hole and set $a = 1.008$. In the near ultracold black hole, the bound is violated when the spin of the black hole is opposite to the angular momentum of a test particle. 

 The right column is for the graphs for a massive particle with $m=1$. We similarly arrange them to the left column. 
 Fig.\ref{fig:Kerr} (b) plots the bound for a massive particle in the flat Kerr black hole. The bound is only violated when the spin parameter is large and opposed to the angular momentum of the test particle. Unlike Fig.\ref{fig:Kerr} (a), the bound is varied when the test particle has a different angular momentum. The Lyapunov exponent increases as the angular momentum of the test particle increases.
 In addition, in the KdS black hole, we can find this property. Fig.\ref{fig:Kerr} (d) shows the bound for the KdS black hole with $\Lambda = 0.1$. The blue region becomes smaller than Fig.\ref{fig:Kerr} (b). The Lyapunov exponent increases as the angular momentum increases when the spin parameter is large.
 Fig.\ref{fig:Kerr} (f) plots the bound for the KNdS black hole with $\Lambda = 0.12$, which includes the near Nariai black hole with $a = 0.56$ and extremal black hole with $a = 1.05$. Fig.\ref{fig:Kerr} (h) plots the bound for the near ultracold black hole with $\Lambda = 0.12$ and $\Lambda = 0.18$. The Lyapunov exponent increases when the angular momentum increases in these graphs. Therefore, in the rotating black hole, the angular momentum makes the motion of the massive particle chaotic. In addition, we can find the same properties that are found in the left column.

 In order to explain why the bound is violated only when the directions of the spin parameter of a black hole and the angular momentum of a test particle are opposite, we consider a rotating particle around a rotating black hole. We consider the distance between a test particle and a fixed point on a black hole. Because the black hole is rotating, the fixed point is moving along the direction of rotation of the black hole. If the direction of the angular momentum is opposite to the spin of the black hole, the distance between the point and test particle is larger than in the case where the directions are the same. Thus, it may appear that the exploring distance is large for the test particle with opposite angular momentum. We conclude that it makes the Lyapunov exponent large and the bound becomes violated when the directions of the spin of a black hole and the angular momentum of a test particle are opposite.

\section{Bound for KNdS black holes}
\begin{figure}[ht]
    \centering
    \includegraphics[width=1\linewidth]{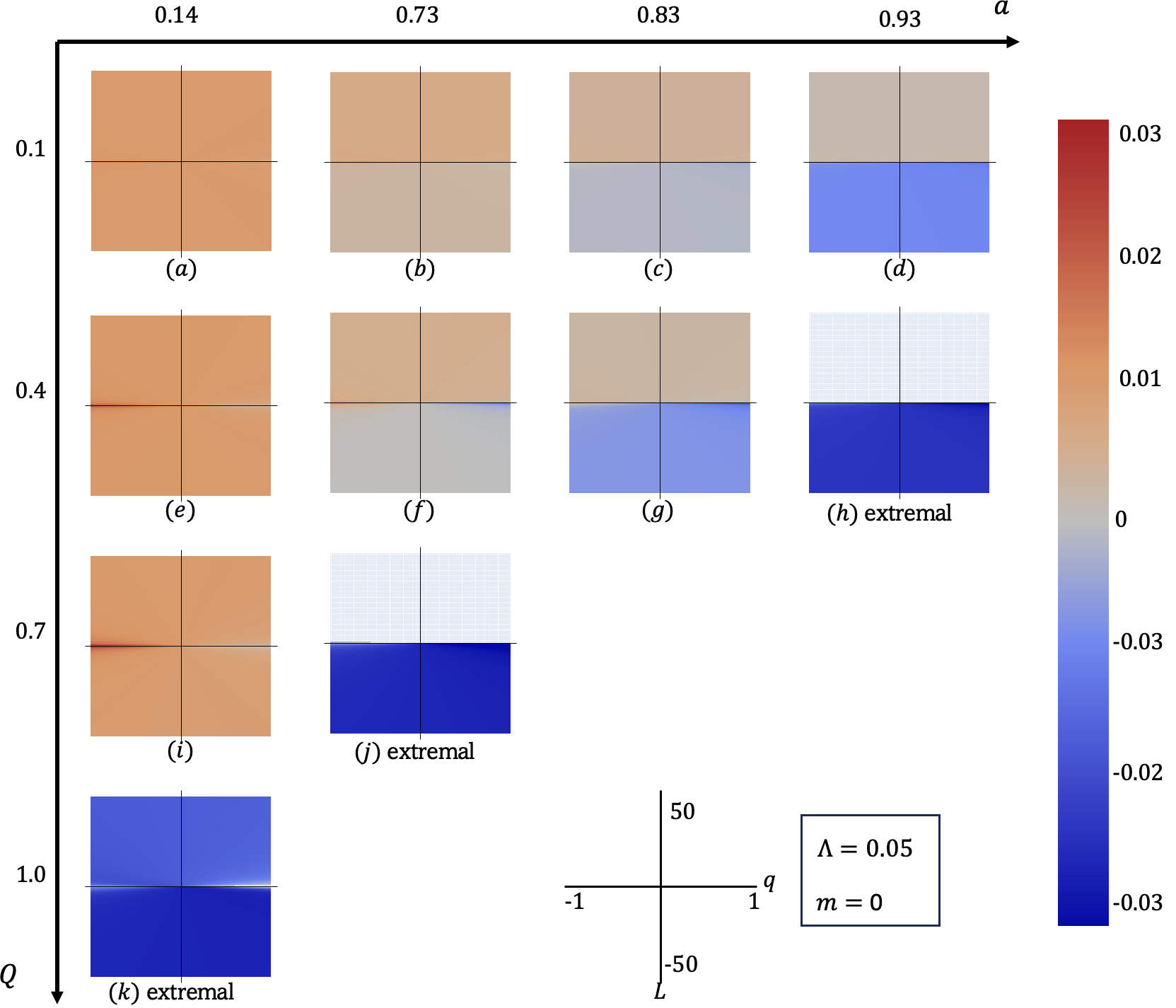}
    \caption{The result for a massless particle in the KNdS black hole with $\Lambda = 0.05$.}
    \label{fig:KN_m=0 0.05}
\end{figure}

\begin{figure}[ht]
    \centering
    \includegraphics[width=1\linewidth]{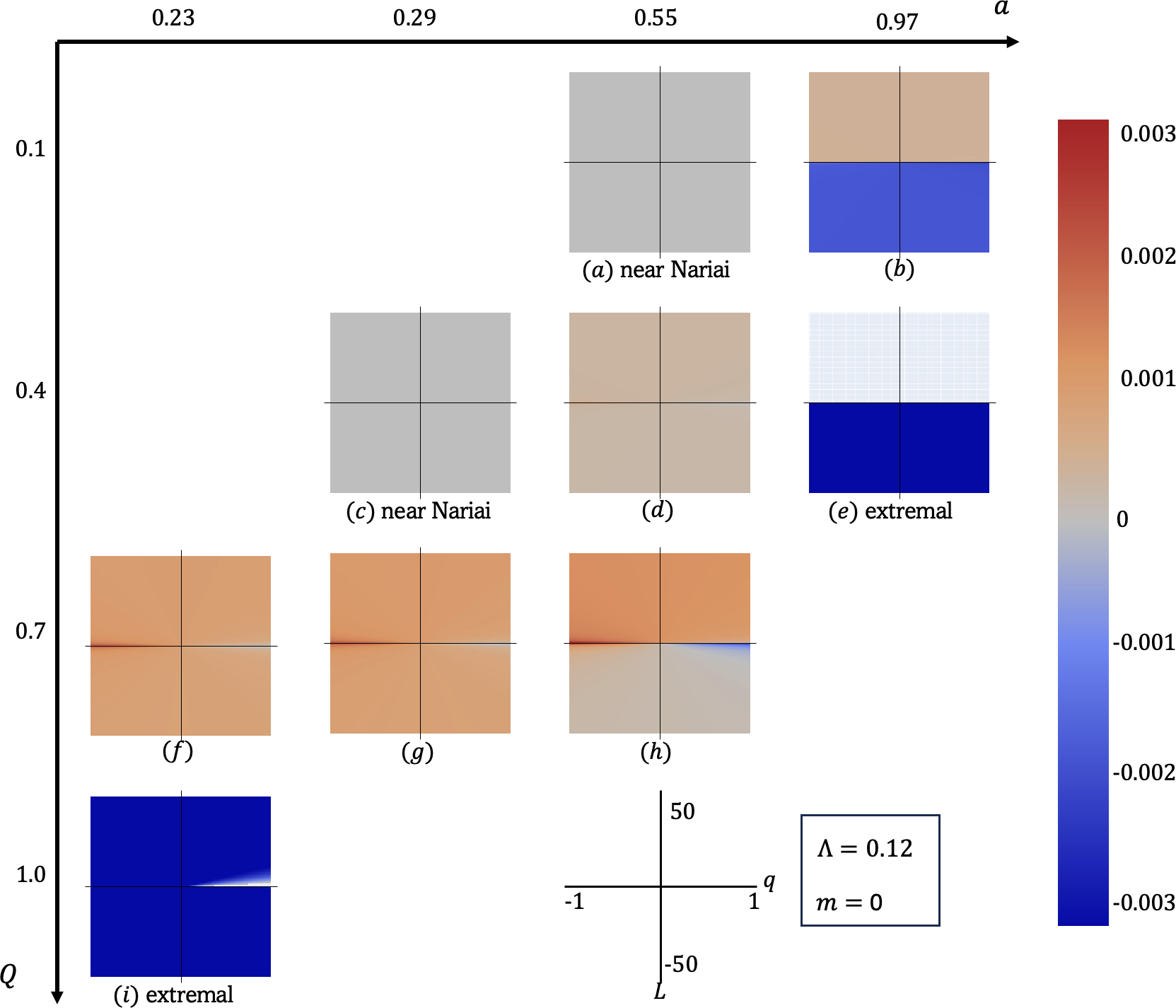}
    \caption{The result for a massless particle in the KNdS black hole with $\Lambda = 0.12$.}
    \label{fig:KN_m=0 0.12}
\end{figure}

\begin{figure}[ht]
    \centering
    \includegraphics[width=1\linewidth]{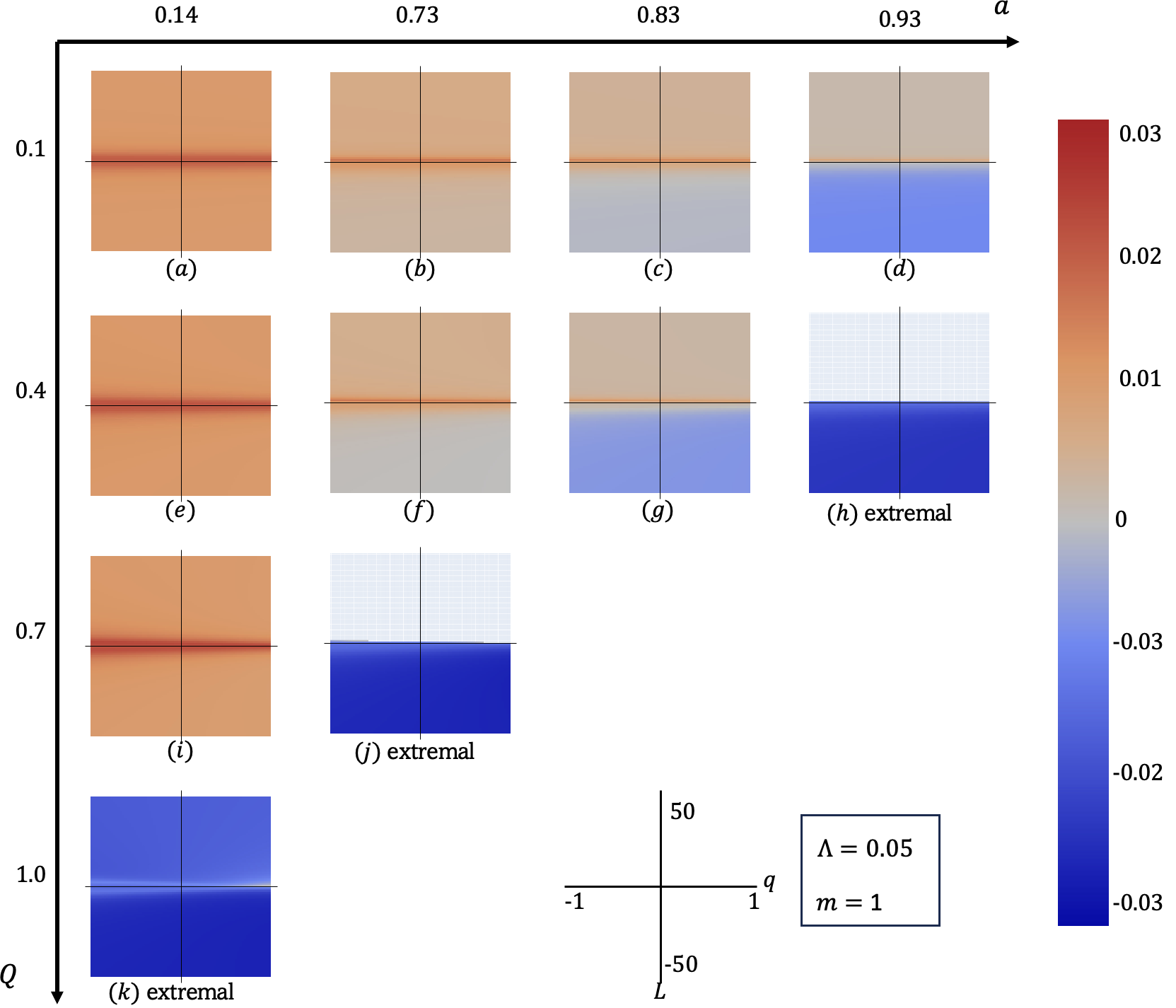}
    \caption{The result for a massive particle with $m=1$ in the KNdS black hole with $\Lambda = 0.05$.}
    \label{fig:KN_m=1 0.05}
\end{figure}

\begin{figure}[ht]
    \centering
    \includegraphics[width=1\linewidth]{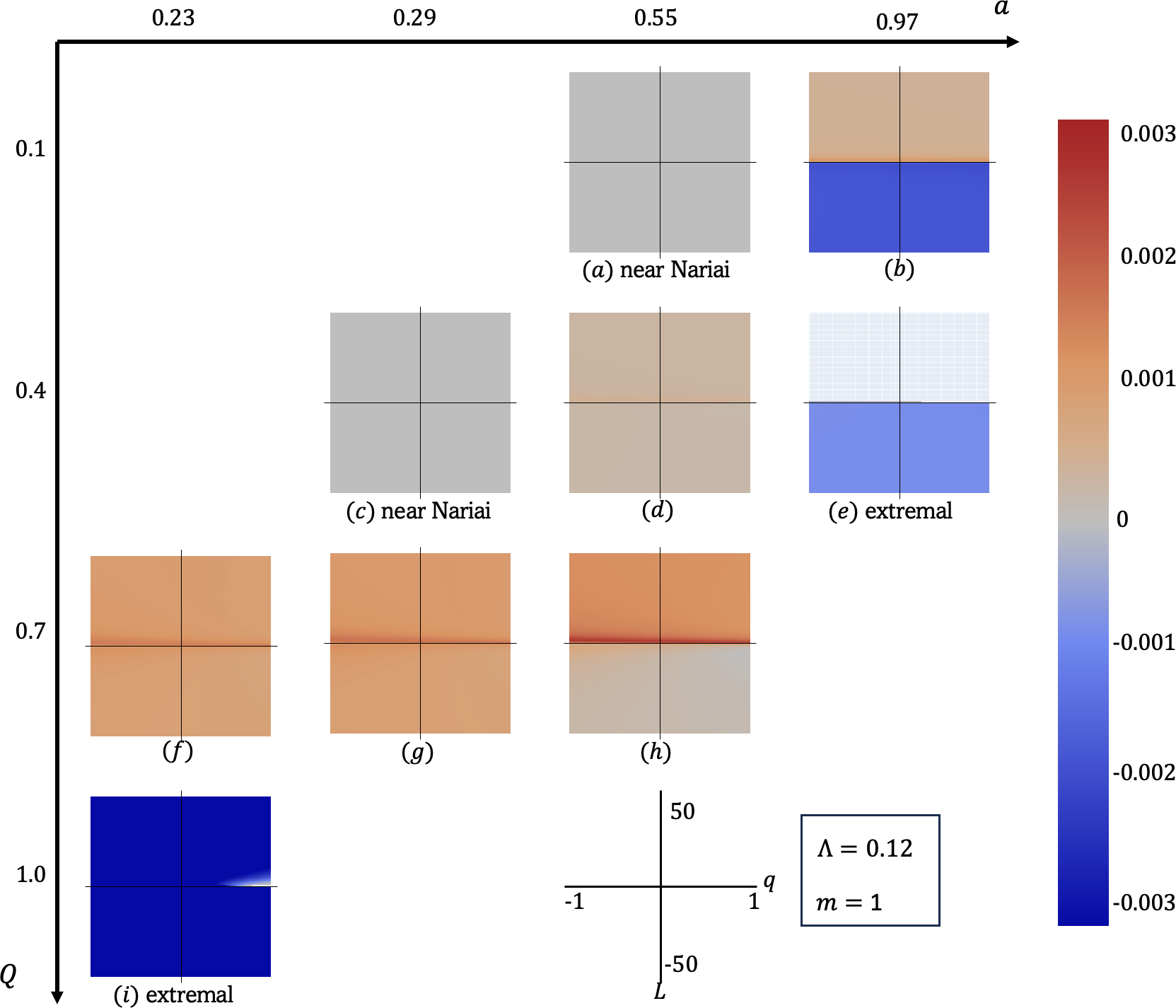}
    \caption{The result for a massive particle with $m=1$ in the KNdS black hole with $\Lambda = 0.12$.}
    \label{fig:KN_m=1 0.12}
\end{figure}

 In this section, we numerically calculate the bound, $\kappa^2 - \lambda^2$, in the KNdS black hole. We plot the graphs for the bound for a massless particle in Fig.\ref{fig:KN_m=0 0.05} and Fig.\ref{fig:KN_m=0 0.12}, and the graphs for the bound for a massive particle in Fig.\ref{fig:KN_m=1 0.05} and Fig.\ref{fig:KN_m=1 0.12}. In addition, we set the cosmological constant as $\Lambda = 0.05$ in Fig.\ref{fig:KN_m=0 0.05} and Fig.\ref{fig:KN_m=1 0.05}, and as $\Lambda = 0.12$ in Fig.\ref{fig:KN_m=0 0.12} and Fig.\ref{fig:KN_m=1 0.12}. Therefore, by comparing these figures, we can determine how the mass of a test particle and the cosmological constant affect the bound (\ref{eq:KNdS exponent}). Further, we set the dimensionless parameter, $M = 1$, in all the figures in this section. Each figure has a vertical and horizontal axis, and we plot the graphs along these axes. We vary the spin parameter of the KNdS black hole on the horizontal axis and the charge parameter of the KNdS black hole on the vertical axis. Therefore, if the graphs lie on the same vertical or horizontal line, the black holes have the same spin parameter or charge parameter. We set the spin parameter and charge parameter to be always positive. In addition, we plot each graph with a vertical and horizontal axis. We vary the electric charge of a test particle on the horizontal axis and the angular momentum of a test particle on the vertical axis. We describe the graphs with color depending on whether the bound is satisfied or not. The blue, red, and white regions indicate that the bound is violated, satisfied, and the motion is not chaotic, respectively. If the graph is for the extremal case, we denote it under the graph. The graphs in the same figure have the same color scale, and we can check it on the right side of each figure.
 
 We plot the graphs for the bound, $\kappa^2 - \lambda^2$, for a massless particle in the KNdS black hole with $\Lambda = 0.05$ in Fig.\ref{fig:KN_m=0 0.05}. In Fig.\ref{fig:KN_m=0 0.05}, there is no graph for a near Nariai KNdS black hole. This is because the Nariai black hole can be found only when the cosmological constant is sufficiently large. 
Fig.\ref{fig:KN_m=0 0.05} (a) is the graph for the KNdS black hole with $a = 0.14$ and $Q = 0.1$. Fig.\ref{fig:KN_m=0 0.05} (b) is the graph for the KNdS black hole with $a = 0.73$ and $Q = 0.1$. In these graphs, the bound is always satisfied at all points. When the angular momentum is positive, the Lyapunov exponent is smaller than the case where the angular momentum is negative.
  Fig.\ref{fig:KN_m=0 0.05} (c) is the graph for the KNdS black hole with $a = 0.83$ and $Q = 0.1$. Fig.\ref{fig:KN_m=0 0.05} (d) is the graph for the KNdS black hole with $a = 0.93$ and $Q = 0.1$. In these graphs, the bound is always satisfied when the angular momentum is positive. Conversely, the bound is always violated when the angular momentum is negative. This means that the direction of the angular momentum determines whether the bound is violated or satisfied.
 Fig.\ref{fig:KN_m=0 0.05} (e) is the graph for the KNdS black hole with $a = 0.14$ and $Q = 0.4$. The bound is always satisfied at all points in Fig.\ref{fig:KN_m=0 0.05} (e). In addition, when the charge of a test particle is positive, the Lyapunov exponent decreases as the absolute value of the angular momentum increases. Conversely, when the charge of a test particle is negative, the Lyapunov exponent increases as the absolute value of the angular momentum increases.
  Fig.\ref{fig:KN_m=0 0.05} (f) is the graph for the KNdS black hole with $a = 0.73$ and $Q = 0.4$. There is a small blue region with small negative angular momentum and a large positive electric charge. Except for this region, the bound is always satisfied. When the angular momentum is negative, the Lyapunov exponent is larger than the case where the angular momentum is positive, which is similar to Fig.\ref{fig:KN_m=0 0.05} (b).
Fig.\ref{fig:KN_m=0 0.05} (g) is the graph for the KNdS black hole with $a = 0.83$ and $Q = 0.4$. The bound is always satisfied when the angular momentum is positive. With a small negative angular momentum and large negative charge, the bound is satisfied. Except for this region, the bound is violated when the angular momentum is negative.
  Fig.\ref{fig:KN_m=0 0.05} (h) is the graph for the extremal KNdS black hole with $a = 0.93$ and $Q = 0.4$. The bound is always violated when the angular momentum is negative. When the angular momentum is positive, the motion is not chaotic.
Fig.\ref{fig:KN_m=0 0.05} (i) is the graph for the extremal KNdS black hole with $a = 0.93$ and $Q = 0.4$. Fig.\ref{fig:KN_m=0 0.05} (i) has similar properties to Fig.\ref{fig:KN_m=0 0.05} (e). By comparing Fig.\ref{fig:KN_m=0 0.05} (a), (e), and (i), we can find that as the electric charge parameter of the KNdS black hole increases, the variation of the Lyapunov exponent becomes large.
  Fig.\ref{fig:KN_m=0 0.05} (j) is the graph for the extremal KNdS black hole with $a = 0.73$ and $Q = 0.7$. When the angular momentum is negative, the bound is always violated. With a small positive angular momentum and a large negative charge, the particle takes a chaotic motion and the bound is violated. Except for this region, the motion is not chaotic when the angular momentum is positive.
 Fig.\ref{fig:KN_m=0 0.05} (k) is the graph for the extremal KNdS black hole with $a = 0.14$ and $Q = 1$. There is a white region, where the motion is not chaotic, with a small positive angular momentum and a large positive charge of a test particle. Except for this region, the bound is always violated. The Lyapunov exponent increases as the absolute value of the angular momentum increases. When the angular momentum is negative, the Lyapunov exponent is larger than the case where the angular momentum is positive.

  We plot the graphs for the bound, $\kappa^2 - \lambda^2$, for a massless particle in the KNdS black hole with $\Lambda = 0.12$ in Fig.\ref{fig:KN_m=0 0.12}. The large cosmological constant makes the Nariai black hole with specific parameters. Therefore, we plot two graphs for the near Nariai limit. In addition, the large cosmological constant makes the absolute value of the bound smaller than Fig.\ref{fig:KN_m=0 0.05}. We can check it by comparing the color bar on the right side in Fig.\ref{fig:KN_m=0 0.05} and Fig.\ref{fig:KN_m=0 0.12}.
   Fig.\ref{fig:KN_m=0 0.12} (a) and (c) are the graphs for the near Nariai KNdS black hole. Because the surface gravity at the event horizon is near zero and the Lyapunov exponent is small, the graphs appear to be gray. However, the bound is always satisfied in these two graphs.
 Fig.\ref{fig:KN_m=0 0.12} (b) is the graph for the KNdS black hole with $a = 0.97$ and $Q = 0.1$. In Fig.\ref{fig:KN_m=0 0.12} (b), the direction of the angular momentum determines whether the bound is violated or not.
  Fig.\ref{fig:KN_m=0 0.12} (d) is the graph for the KNdS black hole with $a = 0.55$ and $Q = 0.4$. The bound is always satisfied in Fig.\ref{fig:KN_m=0 0.12} (d). The bound has a near-zero value when the absolute value of the angular momentum is small and the electric charge of a test particle is positive.
In Fig.\ref{fig:KN_m=0 0.12} (e), the direction of the angular momentum determines whether the motion is chaotic or not. This is similar to Fig.\ref{fig:KN_m=0 0.12} (b). Fig.\ref{fig:KN_m=0 0.12} (e) is the graph for the extremal KNdS black hole with $a = 0.97$ and $Q = 0.4$. The bound is always violated when the angular momentum is negative. However, the particle with positive angular momentum always has a non-chaotic motion.
 Fig.\ref{fig:KN_m=0 0.12} (f) is the graph for the KNdS black hole with $a = 0.23$ and $Q = 0.7$. The bound is always satisfied in Fig.\ref{fig:KN_m=0 0.12} (f), which has similar properties as Fig.\ref{fig:KN_m=0 0.05} (a), (e), and (i).
  Fig.\ref{fig:KN_m=0 0.12} (g) is the graph for the KNdS black hole with $a = 0.29$ and $Q = 0.7$. The bound is always satisfied in Fig.\ref{fig:KN_m=0 0.12} (g). The bound has a near-zero value when the absolute value of the angular momentum is small and the electric charge of a test particle is positive.
 Fig.\ref{fig:KN_m=0 0.12} (h) is the graph for the KNdS black hole with $a = 0.55$ and $Q = 0.7$. There is a small blue region with a small negative angular momentum and a large positive charge. Except for this region, the bound is always satisfied.
 Fig.\ref{fig:KN_m=0 0.12} (i) is the graph for the extremal KNdS black hole with $a = 0.23$ and $Q = 1$. There is a small white region with a small positive angular momentum and a large positive electric charge. Except for this region, the bound is always violated.

  We plot the graphs for the bound, $\kappa^2 - \lambda^2$, for a massive particle, $m = 1$, in the KNdS black hole with $\Lambda = 0.05$ in Fig.\ref{fig:KN_m=1 0.05}. The arrangement is the same with Fig.\ref{fig:KN_m=0 0.05}.
We plot the graph for the KNdS black hole with $a = 0.14$ and $Q = 0.1$ in Fig.\ref{fig:KN_m=1 0.05} (a). The bound is always satisfied in Fig.\ref{fig:KN_m=1 0.05} (a), and the Lyapunov exponent increases as the absolute value of the angular momentum increases.
 Fig.\ref{fig:KN_m=1 0.05} (b) is the graph for the KNdS black hole with $a = 0.73$ and $Q = 0.1$. In Fig.\ref{fig:KN_m=1 0.05} (b), the bound is always satisfied. In addition, the bound with a positive angular momentum is larger than the bound with a negative angular momentum.
  Fig.\ref{fig:KN_m=1 0.05} (c) and (d) are the graphs for the KNdS black holes with $Q =0.1$ and for $a = 0.83$ and $a = 0.93$, respectively. These two graphs have similar properties. The bound is satisfied when the angular momentum is positive in these graphs. The bound with a small negative angular momentum is also satisfied. However, except for these, the bound is violated with a negative angular momentum.
 We plot the graph for the KNdS black hole with $a = 0.14$ and $Q = 0.4$ in Fig.\ref{fig:KN_m=1 0.05} (e), which has similar properties to Fig.\ref{fig:KN_m=1 0.05} (a). However, the variation of the bound becomes larger as the charge parameter of the KNdS black hole increases.
  Fig.\ref{fig:KN_m=1 0.05} (f) is the graph for the KNdS black hole with $a = 0.73$ and $Q = 0.4$. Fig.\ref{fig:KN_m=1 0.05} (f) has similar properties to Fig.\ref{fig:KN_m=1 0.05} (b). However, the difference between the bound with a negative and positive angular momentum becomes larger than Fig.\ref{fig:KN_m=1 0.05} (b).
 Fig.\ref{fig:KN_m=1 0.05} (g) is the graph for the KNdS black hole with $Q =0.4$ and $a = 0.83$. Fig.\ref{fig:KN_m=1 0.05} (g) has similar properties to Fig.\ref{fig:KN_m=1 0.05} (c) and (d). We can also find that the narrow red region with a small negative angular momentum becomes narrow as the electric charge of a test particle increases.
  Fig.\ref{fig:KN_m=1 0.05} (h) is the graph for the extremal KNdS black hole with $a = 0.93$ and $Q = 0.4$. The bound is always violated when the angular momentum is negative. When the angular momentum is positive, the motion is not chaotic. These are the same properties as in Fig.\ref{fig:KN_m=0 0.05} (h).
 Fig.\ref{fig:KN_m=1 0.05} (i) is the graph for the extremal KNdS black hole with $a = 0.14$ and $Q = 0.7$. We observe that Fig.\ref{fig:KN_m=1 0.05} (i) appears to be similar to Fig.\ref{fig:KN_m=1 0.05} (a) and (e), and they have the same properties.
  Fig.\ref{fig:KN_m=1 0.05} (j) is the graph for the extremal KNdS black hole with $a = 0.73$ and $Q = 0.7$. When the angular momentum is negative, the bound is always violated. In addition, the bound with a small positive angular momentum is violated.
 Fig.\ref{fig:KN_m=1 0.05} (k) is the graph for the extremal KNdS black hole with $a = 0.14$ and $Q = 1$. The bound is always violated in Fig.\ref{fig:KN_m=1 0.05} (k). The Lyapunov exponent increases as the absolute value of the angular momentum increases. Further, there is a small gray region where the bound is near zero, with a small positive angular momentum and a large positive charge.

 We plot the graphs for the bound, $\kappa^2 - \lambda^2$, for a massive particle, $m = 1$, in the KNdS black hole with $\Lambda = 0.12$ in Fig.\ref{fig:KN_m=1 0.12}. The arrangement is the same as Fig.\ref{fig:KN_m=0 0.12}. The scale of the graphs is different from Fig.\ref{fig:KN_m=1 0.05}, which can be found in the color bar on the right side of Fig.\ref{fig:KN_m=1 0.12}.
 We plot the graphs for the near Nariai KNdS black holes in Fig.\ref{fig:KN_m=1 0.12} (a) and (c). Similar to Fig.\ref{fig:KN_m=0 0.12} (a) and (c), the bound has the near zero value. However, the bound is always satisfied.
Fig.\ref{fig:KN_m=1 0.12} (b) is the graph for the KNdS black hole with $a = 0.97$ and $Q = 0.1$. The direction of the angular momentum determines whether the bound is satisfied or not. Therefore, when the angular momentum is positive, the bound is satisfied, and when the angular momentum is negative, the bound is violated.
 Fig.\ref{fig:KN_m=1 0.12} (d) is the graph for the KNdS black hole with $Q = 0.4$ and $a = 0.55$. The bound is always satisfied in Fig.\ref{fig:KN_m=1 0.12} (d). In addition, the Lyapunov exponent increases as the absolute value of the angular momentum increases.
Fig.\ref{fig:KN_m=1 0.12} (e) is the graph for the extremal KNdS black hole with $a = 0.97$ and $Q=0.4$. When the angular momentum is negative, the bound is always violated. The bound is also violated with a small positive angular momentum. Except for these regions, the motion of a test particle is not chaotic.
   Fig.\ref{fig:KN_m=1 0.12} (f), (g), and (h) are the graphs for the KNdS black holes with $Q = 0.7$, and for $a = 0.23$, $a = 0.29$ and $a = 0.55$, respectively. In these three graphs, the bound is always satisfied. The bound with a negative angular momentum is smaller than with a positive angular momentum. By comparing these three graphs, we can find that the variation of the bound becomes large as the spin parameter increases.
  Fig.\ref{fig:KN_m=1 0.12} (i) is the graph for the extremal KNdS black hole with $a = 0.23$ and $Q = 1$. There is a small white region with a small positive angular momentum and a large positive charge. Except for this region, the bound is always violated.

 There are common properties that we can find by comparing the graphs. When we compare the graphs in the same horizontal axis, which are for the KNdS black holes that have the same electric charge parameter, the bound with negative angular momentum becomes violated as the spin parameter increases. This means that when the spin parameter is large, the direction of the angular momentum determines whether the bound is satisfied or not. Except for the graphs for extremal black holes with a low spin parameter, the bound with a negative angular momentum tends to become violated as the spin parameter increases. This is similar to the graphs for the KdS black holes. In addition, by comparing the graphs in the same vertical axis, which are for the KNdS black holes that have the same spin parameter, the variation of the bound becomes large as the charge parameter of the black hole increases. When the KNdS black hole has a low spin parameter, the graphs look nearly symmetric. This property is similar to the graphs for the RNdS black holes. When we compare the figures with a massless and massive particle, we can find that the bound tends to be more satisfied when a test particle is massive. The Lyapunov exponent also tends to decrease. Thus, the mass of a test particle makes the motion less chaotic.

   \begin{figure}[ht]
    \begin{subfigure}{0.45\textwidth}
        \centering
        \includegraphics[width=1.1\linewidth]{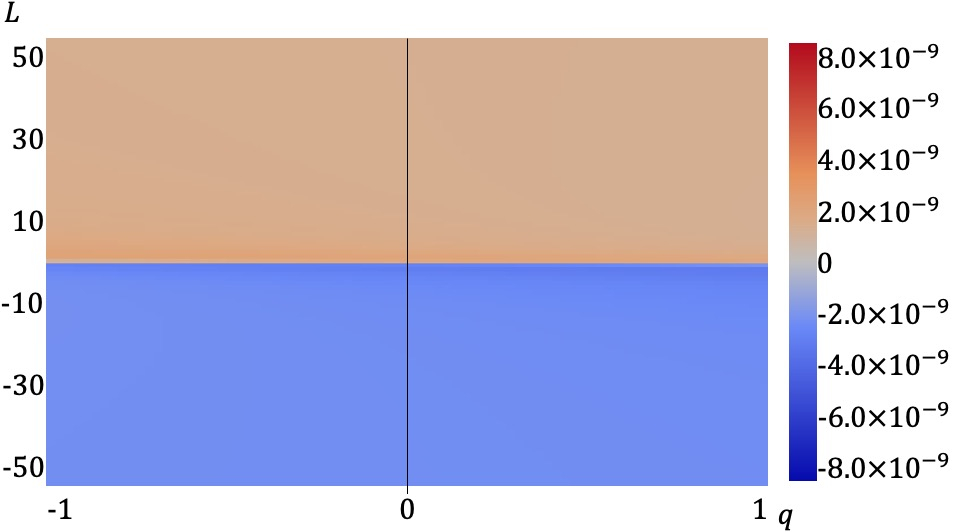}
        \caption{near Nariai ($a = Q,\Lambda$) = $(0.71,0.17)$}
        \label{fig:graph_f}
    \end{subfigure}
    \hfill
    \begin{subfigure}{0.45\textwidth}
        \centering
        \includegraphics[width=1.1\linewidth]{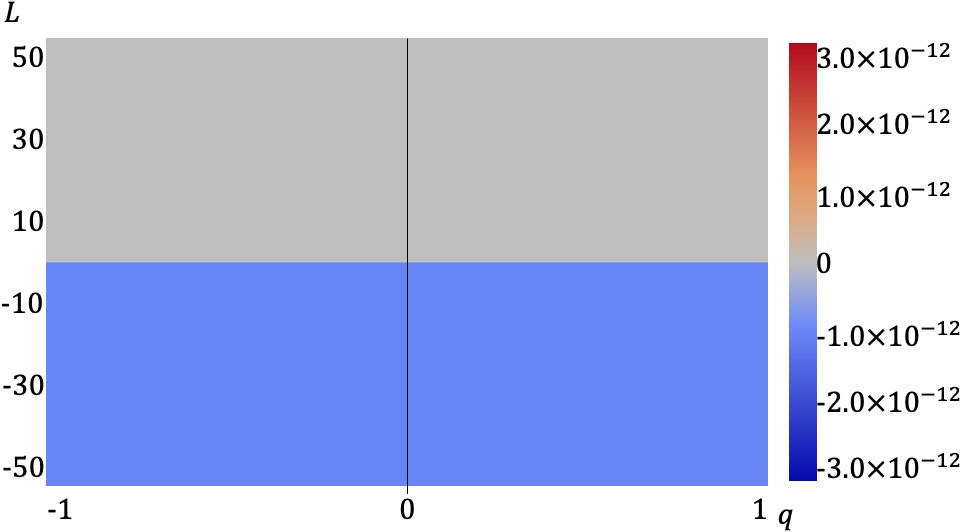}
        \caption{near Ultracold ($a = Q,\Lambda$) = $(0.76,0.1975)$}
        \label{fig:graph_h}
    \end{subfigure}
    \caption{We plot the bound for the near Narai KNdS black hole in (a). We set the parameter as $a = Q = 0.71$ and $\Lambda = 0.17$. We plot the bound for the near ultracold KNdS black hole in (b). The parameters of the near ultracold KNdS black hole are $a = Q = 0.76$ and $Q = 0.1975$. In these two graphs, we consider that a test particle is massive, $m=1$, and $M =1$.}
    \label{fig:KN limit}
\end{figure}
  In the graphs for the near Nariai KNdS black hole, Fig.\ref{fig:KN_m=0 0.12} (a), (c), Fig.\ref{fig:KN_m=1 0.12} (a), and (c), the bound is always satisfied. However, it does not mean that the bound is always satisfied when the black hole is the near Nariai limit. Fig.\ref{fig:KN limit} (a) is the graph for the near Nariai KNdS black hole with $a = Q = 0.71$ and $\Lambda = 0.17$. Fig.\ref{fig:KN limit} (b) is the graph for the near ultracold KNdS black hole with $a = Q = 0.76$ and $Q = 0.1975$. We can find that the bound is violated when an angular momentum is negative in these graphs. The bound is satisfied with a positive angular momentum. Therefore, the fact that it is the near Nariai black hole does not assure that the bound is satisfied. 

Moreover, we can compare the results in this section and the analytical calculation. The graph for the KNdS black hole, which has the smallest electric charge and spin parameters, is Fig.\ref{fig:KN_m=0 0.05} (a). According to Fig.\ref{fig:KN_m=0 0.05} (a), the bound is always satisfied. In addition, the average of the extremum point $r_0$ in Fig.\ref{fig:KN_m=0 0.05} (a) is $r \approx 2.99$. According to Section 3.2, the extremum point is determined by the mass parameter of a black hole in the low spin-charge limit, $r_0 = 3M$. We set the mass parameter as $M = 1$ in the numerical calculation. Thus, Fig.\ref{fig:KN_m=0 0.05} (a) can be matched with the low spin-charge limit, where the bound is always satisfied and a test particle is massless. We aim to consider the results of the numerical methods with the near-horizon limit and near cosmological horizon limit. However, there is no numerical result where the extremum point $r_0$ is not close enough to the event horizon or cosmological horizon to call the near-horizon limit or near cosmological horizon limit.

\section{Summary}
  This study investigated whether the bound (\ref{eq:bh conjecture}) is satisfied or violated in KNdS black holes when a test particle has an electric charge and angular momentum. Using the Polyakov-type action, we calculate the maximum Lyapunov exponent of a test particle. Then, we used analytical and numerical methods.

  Using the analytic method, we calculated the Lyapunov exponent for three limits, the low spin-charge limit, near-horizon limit, and near cosmological horizon limit. In the low spin-charge limit, we assume that the black hole has a low electric charge and spin parameters. With the low spin-charge limit, the mass parameter of the black hole determines the Lyapunov exponent of a test particle. We find that with the low spin-charge limit, the bound is always satisfied.
  In the near-horizon limit and near cosmological horizon limit, specific variables make these limits. We just assume the situation where the extremum point of the effective potential $r_0$ is near the outer horizon and cosmological horizon. With the near-horizon limit, the bound is always violated when the cosmological constant is positive. With the near cosmological horizon limit, the bound is dependent on the surface gravity at the outer horizon and cosmological horizon. Therefore, we cannot simply determine whether the bound is satisfied or not in the near cosmological horizon limit. Considering the near-horizon limit and near cosmological horizon limit, we can find that the bound does not linearly depend on the distance between the extremum point $r_0$ and outer horizon $r_+$. To continue the study, we choose to calculate the bound by using the numerical method in various cases.

   Using the numerical methods, we investigate the bound in three types of black holes, i.e., RNdS, KdS, and KNdS black holes. From the results of the numerical method, the bound is determined by a complex combination of seven variables, $a, M, Q, q, L, m$, and $\Lambda$.
   We plot the graphs for the RNdS black holes in Fig.\ref{fig:m=0_RN} and Fig.\ref{fig:m=1_RN}. For the RNdS black holes, $a \to 0$, the sign of the electric charge of the test particle plays an important role in determining the bound. There is a non-linear relation between the Lyapunov exponent and the angular momentum.
   We plot the graphs for the KdS cases in Fig.\ref{fig:Kerr}. For the KdS black holes, $Q \to 0$, we do not need to consider the electric force. The directions of the spin parameter and angular momentum of a test particle are important to determine the bound. When $m=0$, the bound is the same although the angular momentum becomes different. For both cases, $m=0$ and $m=1$, the bound is violated only when the directions are opposite in the KdS black hole. Varying the angular momentum does not have a large effect on the bound.
   We plot the graphs for the KNdS black hole in Fig.\ref{fig:KN_m=0 0.05}, Fig.\ref{fig:KN_m=0 0.12}, Fig.\ref{fig:KN_m=1 0.05}, and Fig.\ref{fig:KN_m=1 0.12}. For the KNdS black holes, we can find properties similar to the RNdS and KdS black holes. When the electric charge and spin parameters of the KNdS black hole are sufficiently large, we can find that the bound is violated. In particular, the directions of the spin of the black hole and angular momentum of a test particle are opposed; the bound is violated more easily than in the cases where the directions are the same.
  These results can be compared with the result in \cite{Hashimoto:2016dfz, Gwak:2022cha} which investigated test particles with no angular momentum and for the negative cosmological constant.
  
   We can compare the results of analytic calculation with low spin-charge and numerical calculations. The low spin-charge limit agrees well with the numerical results. However, in the numerical calculation, we did not find cases that we can call the near-horizon limit or near cosmological horizon limit. The extremum point $r_0$ is non-linearly dependent on the seven variables, $a, M, Q, q, L, m$, and $\Lambda$. Therefore, we cannot directly compare the results from the numerical calculation with the near-horizon limit and near cosmological horizon limit.
   In this study, we research the bound, $\kappa^2 - \lambda^2$, for the KNdS black hole with a test particle that has an electric charge and angular momentum. Based on the results, we found that the bound is violated in KNdS black holes.

\vspace{0.1in}

\noindent{\bf Acknowledgments}

{\small\noindent This research was supported by Basic Science Research Program through the National Research Foundation of Korea (NRF) funded by the Ministry of Education (NRF-2022R1I1A2063176) and the Dongguk University Research Fund of 2023. BG appreciates APCTP for its hospitality during completion of this work.\\}

\bibliographystyle{jhep}
\bibliography{ref_v2}
\end{document}